\documentclass[twocolumn,english,superscriptaddress,amsmath,floatfix,longbibliography,floats,prx]{revtex4-1}
\usepackage[colorlinks=true,urlcolor=blue,citecolor=blue,linkcolor=blue]{hyperref} 
\usepackage[T1]{fontenc}
\usepackage[utf8]{inputenc}
\usepackage{amssymb}
\usepackage{graphicx}
\usepackage{amsmath,color}
\usepackage{mathrsfs}
\usepackage{float}
\usepackage{indentfirst}
\usepackage{txfonts}
\usepackage[normalem]{ulem}
\usepackage[table]{xcolor}
\usepackage{array,ragged2e}



\usepackage{babel}

\begin{document}

\hyphenpenalty=5000

\tolerance=1000

\title{Gapless quantum spin liquid and global phase diagram of the  spin-1/2 $J_1$-$J_2$  square antiferromagnetic Heisenberg  model}
\author{Wen-Yuan Liu}
 \affiliation{Department of Physics, The Chinese University of Hong Kong, Shatin, New Territories, Hong Kong, China}
 \author{Shou-Shu Gong}
 \affiliation{Department of Physics, Beihang University, Beijing 100191, China}
  \author{Yu-Bin Li}
 \affiliation{Department of Physics, The Chinese University of Hong Kong, Shatin, New Territories, Hong Kong, China}
\author{Didier Poilblanc}
\affiliation{Laboratoire de Physique Th\'eorique, C.N.R.S. and Universit\'e de Toulouse, 31062 Toulouse, France}
 \author{Wei-Qiang Chen}
 \affiliation{Institute for Quantum Science and Engineering and Department of Physics, Southern University of Science and Technology, Shenzhen 518055, China}
\affiliation{Shenzhen Key Laboratory of Advanced Quantum Functional Materials and Devices, Southern University of Science and Technology, Shenzhen 518055, China}
 \author{Zheng-Cheng Gu}
 \affiliation{Department of Physics, The Chinese University of Hong Kong, Shatin, New Territories, Hong Kong, China}

\date{\today }

\begin{abstract}

The nature of the zero-temperature phase diagram of the spin-$1/2$ $J_1$-$J_2$ Heisenberg model on a square lattice has been debated in the past three decades, which may hold the key to understand high temperature superconductivity. By using the state-of-the-art tensor network method, specifically, the finite projected entangled pair state (PEPS) algorithm, to simulate the global phase diagram the  $J_1$-$J_2$ Heisenberg model up to $24\times 24$ sites, we provide very solid evidences to show that the nature of the intermediate nonmagnetic phase is a gapless quantum spin liquid (QSL), whose spin-spin  and dimer-dimer correlations both decay with a power law behavior. There also exists a valence-bond solid (VBS) phase in a very narrow region $0.56\lesssim J_2/J_1\leq0.61$ before the system enters the well known collinear antiferromagnetic phase. The physical nature of the discovered gapless QSL and potential experimental implications are also addressed. We stress that we make the first detailed comparison between the results of PEPS and the well-established density matrix renormalization group (DMRG) method through one-to-one direct benchmark for small system sizes, and thus give rise to a very solid PEPS calculation beyond DMRG. Our numerical evidences explicitly demonstrate the huge power of PEPS for precisely capturing long-range physcis for highly frustrated systems, and also demonstrate  the finite PEPS method is a very powerful approach to study strongly corrleated quantum many-body problems.

\end{abstract}
\maketitle

\section{Introduction} 
Since the discovery of high-Tc cuprates, people conjectured that a spin-$1/2$ antiferromagnetic Heisenberg model on square lattice with nearest neighbor (NN) couplings $J_1$ and next nearest neighbor (NNN) couplings $J_2$ (known as the square lattice $J_1$-$J_2$ model) would support a quantum spin liquid (QSL) phase, which could serve as the primary low-energy metastable states before the systems enter the superconducting phase \cite{RVB,WWZ8913,RS9173,PatrickRVB,RVBsuper}. 
The Hamiltonian of this
model reads:
\begin{equation}
H=J_1\sum_{\langle i,j \rangle}\mathbf{S_i}\cdot\mathbf{S_j}+J_2\sum_{\langle\langle
i,j\rangle\rangle}\mathbf{S_i}\cdot\mathbf{S_j},\quad
(J_1,J_2>0),\label{model}
\end{equation}
For the
intermediate $J_2$ coupling regime, it is long believed that the quantum fluctuation will destroy
the antiferromagnetic (AFM) long range order before the maximally frustrated point $J_2/J_1=0.5$ of the
classical model and might establish a new paramagnetic phase. The nature
of such a paramagnetic phase is of great interest and it might hold the key mechanism of high-Tc cuprates. In early days of high-Tc research, the square lattice $J_1$-$J_2$ model was thus one of the most important frustrated magnet models, and attracted intense research interest, both theoretically and experimentally. In past three decades tremendous efforts by different kinds of methods have been developed to investigate the intermediate paramagnetic phase~\cite{chandra1988,ed1,ed2,CVB1,ed7,CVB2,ed3,ivanov1992,ed5,ed4,PVB1,CVB3,PVB2,VMC2001,gQSL2,PVB3,sirker2006,schmalfu2006,PVB4,PVB5,PVB6,PVB7,murg2009,beach2009,ed6,PVB8,jiang2012,mezzacapo2012,WangRVB,hu2013,PVB9,YangJ1J2,gong2014,chou2014,morita2015,richter2015,wang2016,poilblanc2017,wang2018,CVB4,liu2018,poilblanc2019}, 
and different candidate ground states were proposed, including a columnar valence-bond solid (CVBS) state~\cite{ed1,ed7,ed3,CVB1,CVB2,CVB3,CVB4}, a plaquette valence-bond solid (PVBS) state~\cite{PVB1,PVB2,PVB3,PVB4,PVB5,PVB6,PVB7,PVB8,PVB9,gong2014,morita2015,wang2018,ferrari2020,nomura2021},  and  a gapless QSL state~\cite{VMC2001,gQSL2,WangRVB,hu2013,YangJ1J2,richter2015,poilblanc2017,liu2018,morita2015,wang2018,ferrari2020,nomura2021}, as well as a gapped QSL state~\cite{jiang2012}. Unfortunately, the physical nature of the paramagnetic phase is still enigmatic.

 \begin{figure}[htbp]
 \centering
 \includegraphics[width=3.4in]{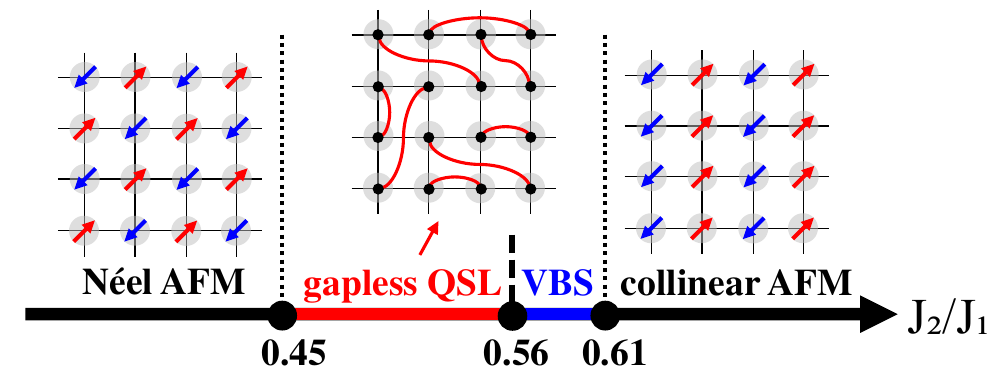}
 \caption{The phase diagram of spin-1/2 $J_1$-$J_2$ square-lattice Heisenberg antiferromagnetic model. The nonmagnetic region is $0.45\lesssim J_2 \leq 0.61$, and it is a gapless spin liquid phase for $0.45\lesssim J_2 \lesssim 0.56$ and a VBS phase for   $0.56 \lesssim J_2 \leq 0.61$.  }
 \label{fig:phaseDiagram}
 \end{figure}

To understand the properties of potential QSL phase of frustrated magnets systematically, X.-G. Wen created the framework of the projective symmetry group (PSG) \cite{W9164} and proposed many different types of QSL variational states for the square lattice $J_1$-$J_2$ model. Among different variational quantum Monte Carlo (vQMC) calculations \cite{VMC2001, hu2013, chou2014, morita2015}, a particular gapless  $Z_2$ QSL state was  intensively studied  with Lanczos projection \cite{hu2013}. As its variational energy is the lowest one among all possible QSL constructed by projective wavefunctions classified by PSG, and is also competitive with the most accurate DMRG ones, people conjecture that such a QSL state could be stabilized in the intermediate paramagnetic phase. However, it is still unclear whether a second order phase transition is possible between such a QSL phase and the usual N\'eel AFM phase. Morevoer, the PSG framework only considers symmetry fractionalization patterns for spinons and, hence,  cannot capture all gapped QSL phases predicted by the general theoretical concept of symmetry enriched topological (SET) order \cite{MengSET}.     Thus, it would not be a surprise if the PSG framework cannot describe all gapless QSL states as well. 

On the other hand, as there are also numerical evidences indicating that a valence-bond solid (VBS) \cite{ed1,ed7,ed3,CVB1,CVB2,PVB1,CVB3,CVB4,PVB2,PVB3,PVB4,PVB5,PVB6,PVB7,PVB8,PVB9,gong2014} might develop in the intermediate paramagnetic phase, an alternative scenario —- the deconfined quantum critical point (DQCP) \cite{DQCP1,DQCP2,sirker2006,DQCP3,DQCP4,DQCP5,DQCP6,DQCP7,DQCP8} was also proposed to describe the direct phase transition between the usual N\'eel AFM phase and the VBS phase. DQCP is an intrinsically strong coupling quantum critical point and it is indeed a Landau forbidden second order phase transition between two ordered phases. This kind of phase transition has already been observed in frustrated-free models, e.g., the $J$-$Q$ model first proposed by Anders Sandvik \cite{DQCP3}.

For convenience, we set $J_1=1$ throughout the whole paper. An early density matrix renormalization group (DMRG) study suggests that the nonmagnetic region $0.41\lesssim J_2\lesssim 0.62$ is a gapped ${\rm Z_2}$ spin liquid phase \cite{jiang2012}, without any spin and dimer orders in the thermodynamic limit. However, a more recent DMRG study with SU(2) symmetry  proposes a PVBS phase for $0.5\lesssim J_2\lesssim 0.61$ with a near critical region $0.44 \lesssim J_2 \lesssim 0.5$ \cite{gong2014}. Later, a very recent DMRG study further proposes two phases in the nonmagnetic region:  a gapless spin liquid phase for $0.46\lesssim J_2\lesssim 0.52$ and a VBS phase for $0.52 \lesssim J_2 \lesssim 0.62$ \cite{wang2018}. On the other hand, a  vQMC  study \cite{hu2013} and a finite projected entangled pair state (PEPS) \cite{liu2018} suggest a gapless QSL phase in the entire intermediate nonmagnetic region. A well known fact is that DMRG is almost numerically exact, but essentially as a one-dimensional algorithm, the precision of DMRG for 2D systems strongly depends on the system width and states kept.  New approaches that can go beyond DMRG for 2D simulation is in great ungency. PEPS, a higher dimensional extension of DMRG, which is  also a systemtically improvable variational ansatz,  provides a very promising tool for solving 2D quantum many-body problems~\cite{verstraete2004,verstraete2006,verstraete2008}.   However, the expensive cost  of PEPS greatly limits its practical application. Recently, in the scheme of combining variational Monte Carlo  method and tensor network states~\cite{sandvik2007,schuch2008,sandvik2008,wang2011,liu2017}, where physical quantities can be evaluated through Monte Carlo sampling and ground states can be obtained by the means of gradient optimization, an accurate PEPS method was established to deal with finite 2D systems on open boundary conditions (OBC)  \cite{liu2019}, making it possible to simulate large systems with very high precision. Particularly, it allows us to compare PEPS and DMRG results directly on the same system, which could be crucial to clarify some long-standing controversial many-body problems. 
     
 \begin{figure}[htbp]
 \centering
 \includegraphics[width=3.4in]{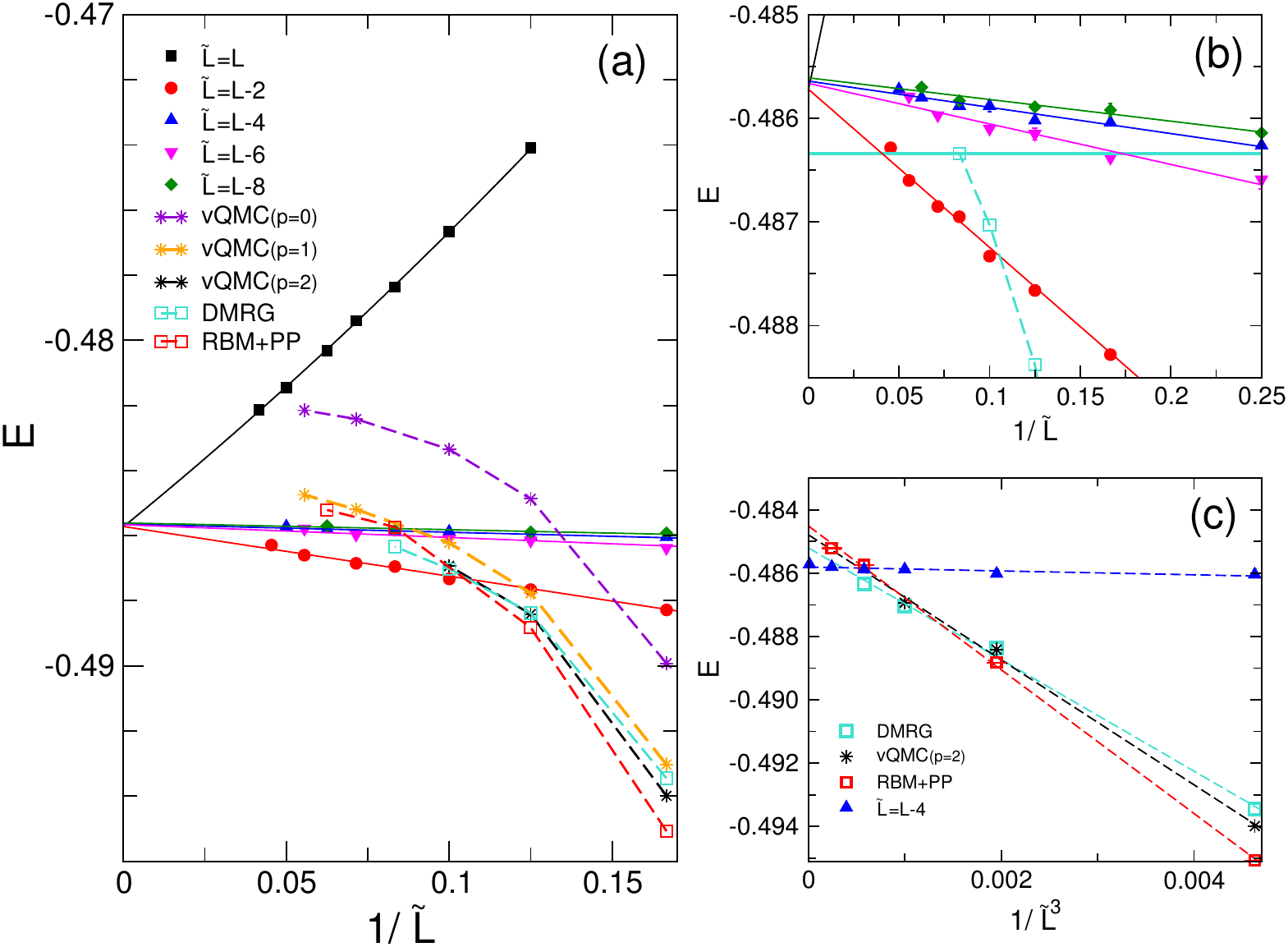}
 \caption{(a) Comparison of ground state energies  in the thermodynamic limit at  $J_2/J_1=0.55$. Different central bulk choices $\tilde{L} \times \tilde{L}$ are used for extrapolations of the $D=8$ optimized PEPS energies. The solid lines denotes  a second order polynomial fit for $\tilde{L}=L$ or a linear fit for the other cases, with extrapolated values lying in a small window about $[-0.4857, -0.4856]$. The vQMC energies with PBC, RBM+PP energies with PBC, and DMRG energies with CBC are presented, taken from Ref.~\cite{hu2013}, Ref.~\cite{nomura2021} and Ref.~\cite{gong2014}, respectively. (b) A detailed scale comparison of finite-size effects between PEPS and DMRG, sharing the same legends with (a). The solid cyan line denotes DMRG energy $-0.4863$ on a cylinder with circumference $L=12$ lattice spacings. (c) Dashed lines denote  fits using $E(\tilde{L})=E_{\infty}+c/\tilde{L}^3$. Energies of $\tilde{L}=L=6-12$ for DMRG ~\cite{gong2014}, $\tilde{L}=L=6-10$ for  vQMC (p=2)~\cite{hu2013}, and  $\tilde{L}=L=6-16$ for RBM+PP~\cite{nomura2021} are used to fit. PEPS energies using $\tilde{L}=L-4$ (blue triangle) are shown for comparison with an extrapolated value $E_{\infty}=-0.48580(4)$ and a coefficient $c=0.061(19)$ via such a linear fit of $1/\tilde{L}^3$.}
 \label{fig:GSenergy_J055}
 \end{figure}
 
      In this paper, we apply the state-of-the-art finite PEPS method to accurately simulate the $J_1$-$J_2$ model up to $24\times 24$. Our results show that the nonmagnetic region $0.45\lesssim J_2\leq 0.61$ consists of a gapless QSL phase for  $0.45\lesssim J_2\lesssim 0.56$ and a VBS phase for  $0.56\lesssim J_2\leq 0.61$. The QSL phase is gapless by observing a power law decay of both spin-spin and dimer-dimer correlation functions. Through detailed comparison with DMRG, we provide very solid numerical results beyond DMRG. We also propose an effective field theory to understand the nature of such a gapless spin liquid and discuss the potential relationship with DQCP scenario. 
      
The rest of the paper is organized as follows. In Sec.{\uppercase\expandafter{\romannumeral2}}, we show the energy comparison with other methods, and present the global phase diagram including critical exponents by measuring order parameters. In Sec.{\uppercase\expandafter{\romannumeral3}}, we compare spin-spin correlations with DMRG calculations and analyze their decay behavior in details.  We also  analyze the decay behavior of dimer-dimer correlations.   In Sec. {\uppercase\expandafter{\romannumeral4}} we discuss the nature of the paramagnetic region and interpret the quantum phase transitions with quantum field theories. In Sec. {\uppercase\expandafter{\romannumeral5}} we discuss our results and explain the origin of different scenarios from other studies.

\bigskip
\section{Global phase diagram}
\subsection{Finite-size scaling of ground state energy.}
We begin with the computation of ground state energies. All energies and order parameters are computed with $D=8$ PEPS, if not otherwise specified. We first compare PEPS ground state energies with available DMRG energies and variational QMC energies. Although our systems are based on open boundaries,  finite-size scaling (FSS) formulas can still work very well~\cite{liu2019}.  In a previous work, it has been shown that our method has extremely high precision for both unfrustated and frustrated models~\cite{liu2019}. For the unfrustrated case, i.e., $J_2=0$, the obtained energies and magnetizations agree excellently with standard (QMC) results. For frustrated cases, the ground state energy in 2D limit at $J_2=0.5$ is $-0.49635(5)$, very close to the corresponding DMRG lower bound energy with $-0.4968$.   Here we check the ground state energy at another highly frustrated point $J_2=0.55$. We compute systems $L\times L$ for $L=8-24$.  Shown in Fig.~\ref{fig:GSenergy_J055}(a),  we use the whole system $\tilde{L}=L$ for FSS versus $1/\tilde{L}$ to obtain the extrapolated energy with $-0.48572(9)$. Alternatively, we can also use other central bulk choices such as $\tilde{L}=L-2$,  $\tilde{L}=L-4$ for FSS, and they give almost the same extrapolated energies. The  DMRG energies with cylindrical boundary condition (CBC) taken from Ref.~\cite{gong2014}, and the vQMC energies with different Lanczos projection steps with periodic boundary condition (PBC)  taken from Ref.~\cite{hu2013} and   are also presented, as well as the PBC energies from a method combining restricted Boltzmann machine and pair-product states (RBM+PP)~\cite{nomura2021}. The DMRG and vQMC p=2 energies on each size look very close, while this might be a coincidence. In Fig.~\ref{fig:GSenergy_J055}(b), we note that the estimated 2D limit DMRG energy $E_{\rm DMRG} \simeq -0.4863$ should be regarded as their lower bound due to finite-size effects. The extrapolated 2D limit energy of PEPS from finite size scaling is about $-0.48572(9)$, very close to the DMRG lower bound energy $-0.4863$. These results are are from different boundary conditions, and we can use their obtained extrapolated energy in the 2D limit for indirect comparison. For frustrated models with CBC and PBC, the precise energy leading scaling with respect to linear system size $L$ is theoretically unclear. Considering CBC and PBC have small boundary effects, usually people can reasonably assume the energy scales as $E(\tilde{L})=E_{\infty}+c/\tilde{L}^3$ where $\tilde{L}=L$ for DMRG, vQMC and RBM+PP results, according to the knowledge on the Heisenberg model~\cite{miles2012}. Such a $1/L^3$ leading scaling indeed seems to work well for fittings, and the obtained 2D energy  of DMRG results using $L=6-12$~\cite{gong2014}, vQMC p=2 results using $L=6-10$~\cite{hu2013} and RBM+PP results using $L=6-16$~\cite{nomura2021} are $-0.4852(2)$, $-0.4848(3)$ and $-0.4845(1)$, respectively, shown in Fig.~\ref{fig:GSenergy_J055}(c). If $L=6$ is excluded, the fits (not shown in Fig.~\ref{fig:GSenergy_J055}(c) for clearness) give corresponding DMRG, vQMC p=2 and RBM+PP extrapolated energy about $-0.48552(7)$, $-0.4854$ (only two points for fit) and $-0.4846(1)$, still higher than our obtained 2D energy $-0.48572(9)$. In anyway, these detailed analyses demonstrate our energy is among the most accurate ones. A similar analysis at $J_2=0.5$ can be found in Appendix.~\ref{app:energyJ2_0.5}. 
 We would mention that because our finite PEPS energies are based on open boundary conditions, due to boundary effects, one should neither directly compare them with DMRG or vQMC results  size by size, nor use the large size energy, say $24\times 24$, as an 2D estimation to directly compare with other 2D estimated energies.  Nevertheless, the central bulk energy like $\tilde{L}=L-4$  show much smaller finite-size effects and can be used to directly estimate the 2D limit energy, as is seen in Fig.~\ref{fig:GSenergy_J055}(a) and (c).

\subsection{Finite-size scaling of order parameters and critical exponents}

\begin{figure}[htbp]
 \centering
 \includegraphics[width=3.4in]{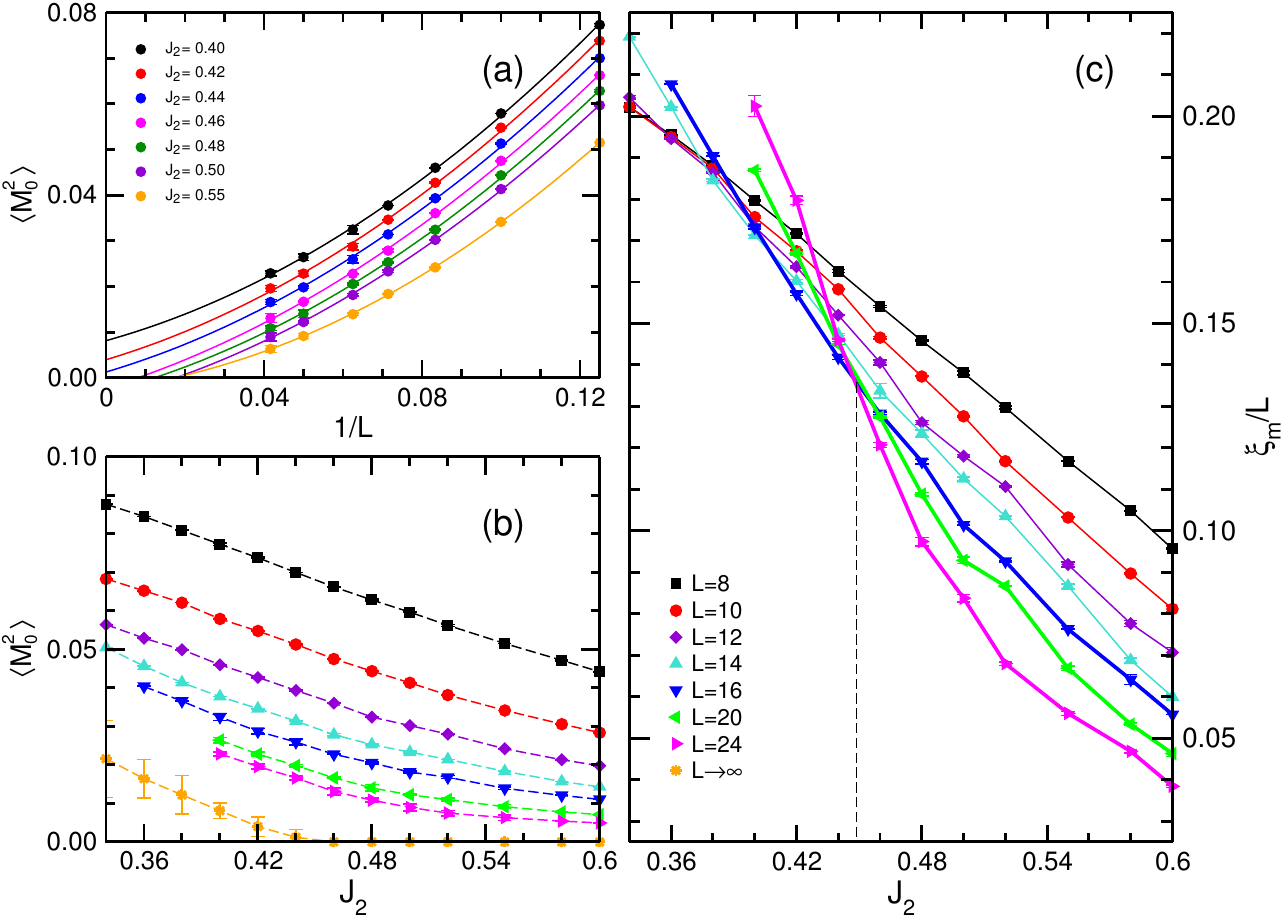}
 \caption{(a) The N\'eel AFM order $\langle {\bf M}_0^2\rangle$ of the $D=8$ optimized PEPS  on $L\times L$ lattices up to $L=24$. Finite-size extrapolations through a second order polynomial fit. (b) The $J_2$ dependence of $\langle {\bf M}_0^2\rangle$ on different systems from $L=8$ to 24. The 2D limit estimated values are also shown as orange sysmbols. (c) The scaling of crossing points of the dimensionless quantity $\xi_m/L$ used to extract critical point $J_{c1}$ at different $J_2$ from $L=8$ to 24.  }
 \label{fig:neelorder}
 \end{figure}

Now we consider spin orders including AFM N\'eel order and collinear order.  The  spin order parameter (squared) is expressed as $m^2_s({\bf k})=\frac{1}{L^4}\sum_{\bf{ij}}\langle{\bf S}_{{\bf i}}\cdot {\bf S}_{{\bf j}}\rangle {e}^{i {\bf k}\cdot({\bf i}-{\bf j})}$, where ${\bf i}=(i_x,i_y)$  is the site position.  
The  N\'eel order parameter (squared) corresponds to the value at  $\bf k_0=(\pi,\pi)$, i.e., $m^2_s({\bf k_0})=\langle {\bf M}_0^2\rangle$,
where 
\begin{equation}
{\bf M}_0=\frac{1}{L^2}\sum_{\bf i} (-1)^{\bf i} {\bf S}_{{\bf i}}\, . 
\end{equation}
In Fig.~\ref{fig:neelorder}(b), we present the AFM N\'eel order on different $L\times L$ systems up to $24\times 24$. Our results suggest that the N\'eel  order vanishes around $J_{c1}= 0.45$ in the thermodynamic limit via FSS.  Note that, for $J_2\ge 0.45$, a power law fit, instead of polynomial one, is more relevant (see Appendix B). Alternatively, we also use a dimensionless quantity $\xi_m/L$ to evaluate the critical point where $\xi_m$ is a correlation length defined as $\xi_m=\frac{L}{2\pi}\sqrt{\frac{m^2_s(\pi,\pi)}{m^2_s(\pi,\pi+2\pi/L)}-1}$~\cite{FSS1}. From the results of $L=16$, 20 and 24 in Fig.~\ref{fig:neelorder}(c), we can see the critical point is indeed located at $J_{c1}\simeq 0.45$, well consistent with the above result. We also find that the collinear AFM order appears at  $J_{c3}=0.61$ via a first order transition, and we will discuss more details for $J_{c3}$ in the Appendix.~\ref{app:stripe}.

 \begin{figure}[htbp]
 \centering
 \includegraphics[width=3.4in]{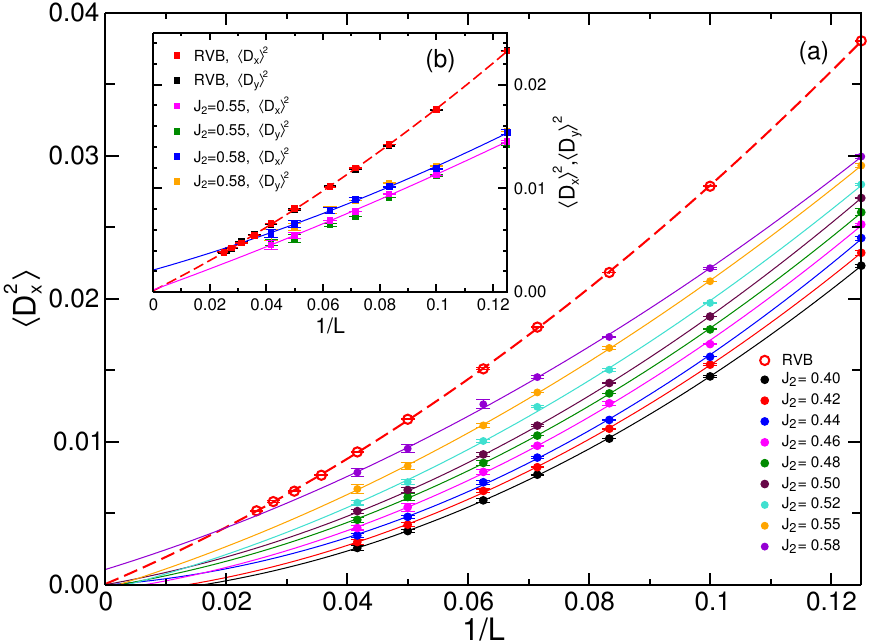}
 \caption{Finite-size scaling of the VBS order parameter  of the $D=8$ optimized PEPS. Results for the RVB state on finite systems are used as references. (a) VBS order parameter based on bond-bond correlations. (b) Local VBS order induced by boundaries. All extrapolations are performed through a second-order polynomial fitting.}
 \label{fig:BondVBSorder}
 \end{figure}

Next we measure the dimer order parameters to detect the possibility of VBS order in the nonmagnetic region $0.45\lesssim J_2 \leq 0.61$. The bond operator is defined as 
$
B^{\alpha}_{\bf i}={\bf S}_{{\bf i}} \cdot {\bf S}_{{\bf i}+{\rm e_\alpha}}
$
between site ${\bf i}$ and site ${\bf i}+{\rm e_\alpha}$ along $\alpha$ direction  with $\alpha=x$ or $y$. Then we can use the dimer order parameter (DOP)
\begin{equation}
D_{\alpha}=\frac{1}{N_b}\sum_{{\bf i}}(-1)^{i_{\alpha}}B^{\alpha}_{\bf i}
\end{equation} 
to detect possible VBS patterns~\cite{zhao2020}, where $N_b=L(L-1)$ is the total number of counted bonds. For spin liquid states, $\langle D^2_\alpha\rangle$  and  $\langle D_{\alpha}\rangle^2$ should be zero in the thermodynamic limit.  In Fig.~\ref{fig:BondVBSorder}(a), we present the horizontal dimer orders $\langle D^2_x\rangle$ for  $J_2$ ranging from 0.40 to 0.58. By extrapolation to 2D limit, we find that the dimer values are zero for $J_2\leq$ 0.55, while there is a small nonzero value for $J_2=0.58$, indicating a potential weak VBS phase. For further analysis, we compare the $J_1$-$J_2$ results with a  resonant valence bond (RVB) state, which is a gapless spin liquid described by a $D=3$ PEPS~\cite{WangRVB}. The dimer orders of such a RVB state are accurately computed on open boundary conditions up to $40\times40$ sites.  Typically, the dimer values of $\langle D^2_x\rangle$ of  $J_1$-$J_2$ model are smaller than those of the RVB state.  For comparison, we consider the finite-size scaling behavior $\langle D^2_x\rangle \sim L^{-(1+\eta'_d)}$ which is expected to be more relevant than the polynomial fit of Fig.~\ref{fig:BondVBSorder}(a) when $\langle D_\alpha\rangle =0$ (see Appendix B). With increasing $J_2$ from 0.40 to 0.55, $\eta'_d$ will decrease from 0.96(2) to 0.32(2), but all of them are larger than  those of the RVB state with $\eta'_d=0.23(1)$, indicating the corresponding extrapolated dimer values must be zero. However, for $J_2=0.58$,  $\eta'_d=0.18(2)$, which is slightly smaller than the RVB state. 

We further directly check the dimer values $\langle D_{\alpha}\rangle^2$ which are induced by the boundaries, shown in Fig.~\ref{fig:BondVBSorder}(b). In general, we find $\langle D_x\rangle^2$ is almost the same as $\langle D_y\rangle^2$, showing a good isotropic behavior between the $x-$ and $y-$ directions. For $J_2=0.55$, the induced dimerizations $\langle D_{\alpha}\rangle^2$ are also zero in the thermodynamic limit, while for $J_2=0.58$  the extrapolated value $\langle D_{\alpha}\rangle^2\simeq 0.0026(11)$ is finite, consistent with $\langle D_{\alpha}^2\rangle\simeq 0.0018(9)$ and, hence, with a VBS phase at $J_2=0.58$. In addition, to further check the extrapolated value  $\langle D^2_x \rangle$ at $J_2=0.58$, we also try second order fittings with different fitting intervals. The extrapolated values by using 4 large-L data, 5 large-L data and 6 large-L data to fit , are  $-0.0007(31)$,  $-0.0009(23)$ and 0.0025(11), respectively.  Note for the 4 large-L data and 5 large-L data fits, the extrapolated values have a large uncertainty, which is probably due to the lack of data to fit.  A third order fitting function for the all available points gives 0.0031(12), consistent with the second-order fitted value 0.0018(9). These extrapolated values are compatible with the induced dimerization $\langle D_{\alpha}\rangle^2\simeq 0.0026(11)$. 
Actually, the VBS is  further supported by the decay behaviours of correlation functions, which we will discuss in next section.  In summary, our results suggest that in the nonmagnetic region $0.45\lesssim J_2 \leq 0.61$, a spin liquid phase covers a large region, and  there is also a small window for a weak VBS phase around $J_2=0.58$.
 
\begin{figure}[htbp]
 \centering
 \includegraphics[width=3.4in]{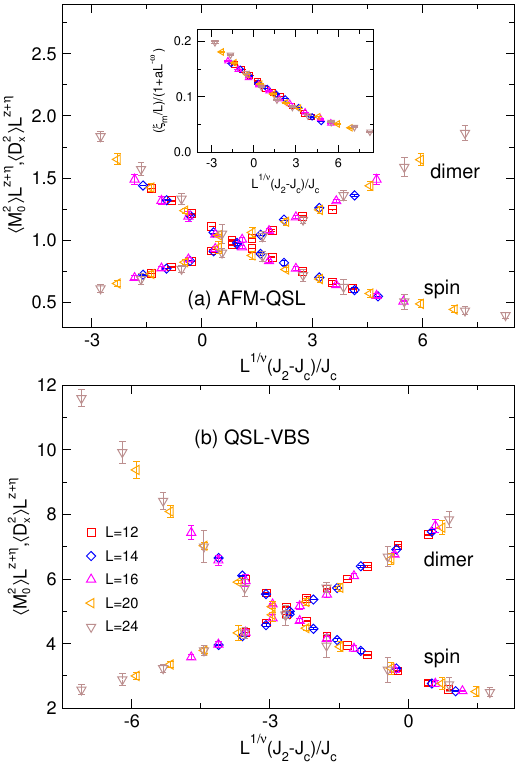}
 \caption { (a) Data collapse of spin and dimer orders for the AFM-QSL transition assuming the critical point located at $J_{c1}=0.45(1)$. (b) Data collapse for the QSL-VBS transition at the critical point  $J_{c2}=0.56(1)$. In case (b), dimer order values are magnified by a factor of 20 for a convenient comparison. The inset presents the scaling of the correlation length $\xi_m$ at $J_{c1}=0.45$ with $\nu=0.99$, $\omega=2$ and $a=14$.}
 \label{fig:dataCollapse}
 \end{figure}

In order to  extract critical exponents,  we further analyse the scaling of obtained quantities. As mentioned above, the AFM-QSL transition point can be  precisely located by the crossing of the dimensionless quantity $\xi_m/L$ at $J_{c1}=0.45(1)$ which has relatively small finite size effects~\cite{FSS1}. In fact, such an approach is equivalent  to the so-called  correlation ratio method~\cite{correlationratio,nomura2021}, both using the value of $m^2_s(\pi,\pi)/m^2_s(\pi,\pi+2\pi/L)$.  In terms of $\xi_m/L$, it can be further used to extract the correlation length exponent $\nu$~\cite{FSS1}. For the QSL-VBS transition, in principle similar correlation length from dimer structure factor information can be also applied to locate the transition point. Unfortunately, the dimer structure factor is not well defined on open boundary conditions~\cite{zhao2020}. Nevertheless, the QSL-VBS transition point can  still be  located at $J_{c1}=0.56(1)$, according to the finite size scaling of order parameters, as well as the spin-spin correlation decay behaviours which is shown in next part.  Now we extract the correlation length exponents $\nu$ by scaling $\xi_m$ at the AFM-QSL transition point. To obtain a  good data collpase, we use the scaling formula with a subleading correction~\cite{DQCP3}:
\begin{eqnarray}
\xi_m(J_2,L)/L=(1+aL^{-\omega})F_{\xi}[L^{1/\nu}(J_2-J_c)/J_c] 
\end{eqnarray}
As seen in the inset of Fig.~\ref{fig:dataCollapse}(a), the correlation length $\xi_m$ can be scaled with $\nu=0.99(6)$ and  $J_{c1}=0.45(1)$, with the subleading terms  $\omega\approx 2$ and $a\approx 14$. Next we scale  spin and dimer order parameters according to the standard formula~\cite{DQCP3,DQCP7} (here we find subleading corrections are unnecessary):
\begin{eqnarray}
\langle {\bf M}^2_0 \rangle L^{z+\eta_s}=F_s[L^{1/\nu}(J_2-J_c)/J_c] \\
\langle D^2_x \rangle L^{z+\eta_d}=F_d[L^{1/\nu}(J_2-J_c)/J_c]
\end{eqnarray}
where $z$ is the dynamic exponent,  $\eta_s$ and $\eta_d$ are  critical exponents which govern spin and dimer correlations, respectively. $\nu$ is the  correlation length exponent, which can be the same for spin and dimer in the theory of DQCP.  Here we find the obtained $\nu=0.99(6)$  works  well for both spin and dimer cases at the AFM-QSL and QSL-VBS transition points. In Fig.~\ref{fig:dataCollapse} we present data collapse for spin and dimer order parameters using  available points  from $J_2=0.40$ to 0.60 up to $24\times 24$.  Assuming $z=1$ , critical exponents $\eta $ are listed in Table.\ref{tab:exponents}. Comparing the exponents at $J_{c1}$ and $J_{c2}$, note $\eta_{s1}<\eta_{s2}$ and $\eta_{d1}>\eta_{d2}$, consistent with the following results of spin and dimer correlation functions. Actually, from the following parts we know, the QSL is gapless with spin-spin and dimer-dimer correlations both decaying as a power law. A single $\nu=0.99(6)$ seems compatible with such a critical property of the QSL.

In Table~\ref{tab:exponents}. we compare the AFM-QSL and QSL-VBS critical exponents with those from AFM-VBS transition in $J$-$Q$ models (with similar system size) that can be understood  by the DQCP scenario. We note that the spin exponent $\eta_{s1}=0.38(3)$ at the point $J_{c1}=0.45$ abutting to the N\'eel AFM phase and the dimer exponent $\eta_{d2}=0.26(3)$ at the point $J_{c2}=0.56$ abutting to the VBS phase, are intrinsically close to the corresponding exponents of the $J$-$Q$ model, while the correlation length exponent $\nu=0.99(6)$ is obviously different from the one of the $J$-$Q$ model. This might indicate that the critical point associated to the AFM-VBS transition in the DQCP theory can expand into a QSL phase, and we will also provide a potential quantum field theory understanding later.

  \begin{table}[htbp]
   \centering
 \caption { Critical exponents of $J_1$-$J_2$ model  at AFM-QSL  transition point $J_{c1}=0.45$ and QSL-VBS transition point $J_{c2}=0.56$. Because we can only extract the values of $\eta_s$ and $\eta_d$,  here we assume dynamic exponent $z=1$  to obtain $\eta_s$ and $\eta_d$ for comparison. 
  Critical exponents of AFM-VBS in $J$-$Q_2$ model up to $32\times32$ (case a) \cite{DQCP3} and up to $64\times64$ (case b) \cite{DQCP7} as well as  $J$-$Q_3$ model up to $64\times64$ \cite{DQCP7} are listed for comparision. The correlation length exponent $\nu$ of $J$-$Q$ model up to $448\times 448$ is 0.455(2) ~\cite{DQCP8}.}
	\begin{tabular*}{\hsize}{@{}@{\extracolsep{\fill}}ccccc@{}}
		\hline\hline
	   model &type &    $\eta_s$ & $\eta_d$    & $\nu$   \\ \hline
 		$J_1$-$J_2$ & AFM-QSL  & 0.38(3)   & 0.72(4)  & 0.99(6)  \\
 		$J_1$-$J_2$ &QSL-VBS  &0.96(4)   & 0.26(3)  & 0.99(6)  \\
 		$J$-$Q_2$(a) &AFM-VBS & 0.26(3) & 0.26(3) & 0.78(3) \\
 		$J$-$Q_2$(b) &AFM-VBS & 0.35(2) & 0.20(2) & 0.67(1) \\
 		$J$-$Q_3$ &AFM-VBS & 0.33(2) & 0.20(2) & 0.69(1) \\
 		\hline\hline
	\end{tabular*}
\label{tab:exponents}	
\end{table}

\section{Correlation functions in the quantum spin liquid phase}

\subsection{Spin-spin correlation functions}
\subsubsection{A detailed comparison between DMRG and PEPS results}
To investigate the physical nature of the QSL phase in the maximally frustrated region around $J_2\sim 0.5$, we further compute the spin and dimer correlation functions on a strip $L_y\times L_x$ that is fully open along both $x$ and $y$ directions. We compare the results obtained by DMRG with SU(2) spin rotation symmetry and the finite PEPS ansatz. The correlations are measured along the central line $y=L_y/2$ and the distance of the reference site away from the left edge is chosen to be 3 lattice spacings in order to minimize boundary effects. 

\begin{figure*}[htbp]
 \centering
 \includegraphics[width=6.8in]{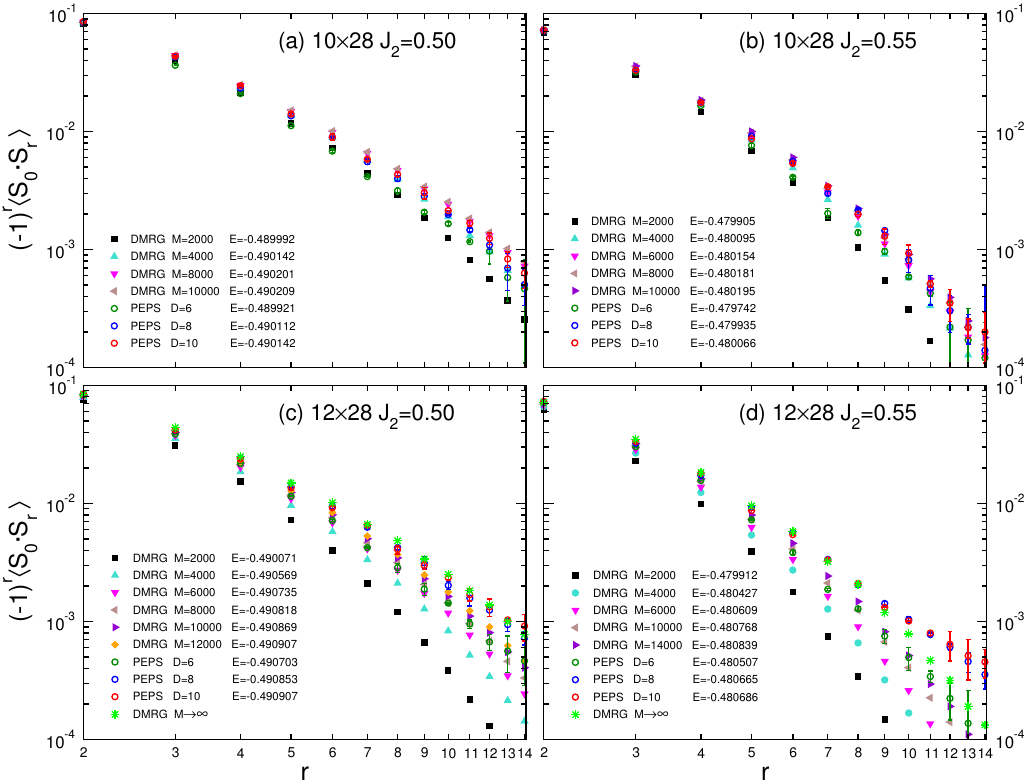}
 \caption{Comparison of spin-spin correlations vs distance obtained in DMRG and PEPS calculations for various bond dimensions $M$ and $D$.  $10\times 28$ and $12\times28$ systems at $J_2=0.5$ and $0.55$ are considered in the four panels (a-d). Energies for the different $M$ or $D$ are listed in the corresponding legends for comparison. A log-log plot is used so that the expected power-law behavior in the 2D limit (see text) would show up as a straight line.}
 \label{fig:AllSpinCorr}
 \end{figure*}

  \begin{figure}[htbp]
 \centering
 \includegraphics[width=3.4in]{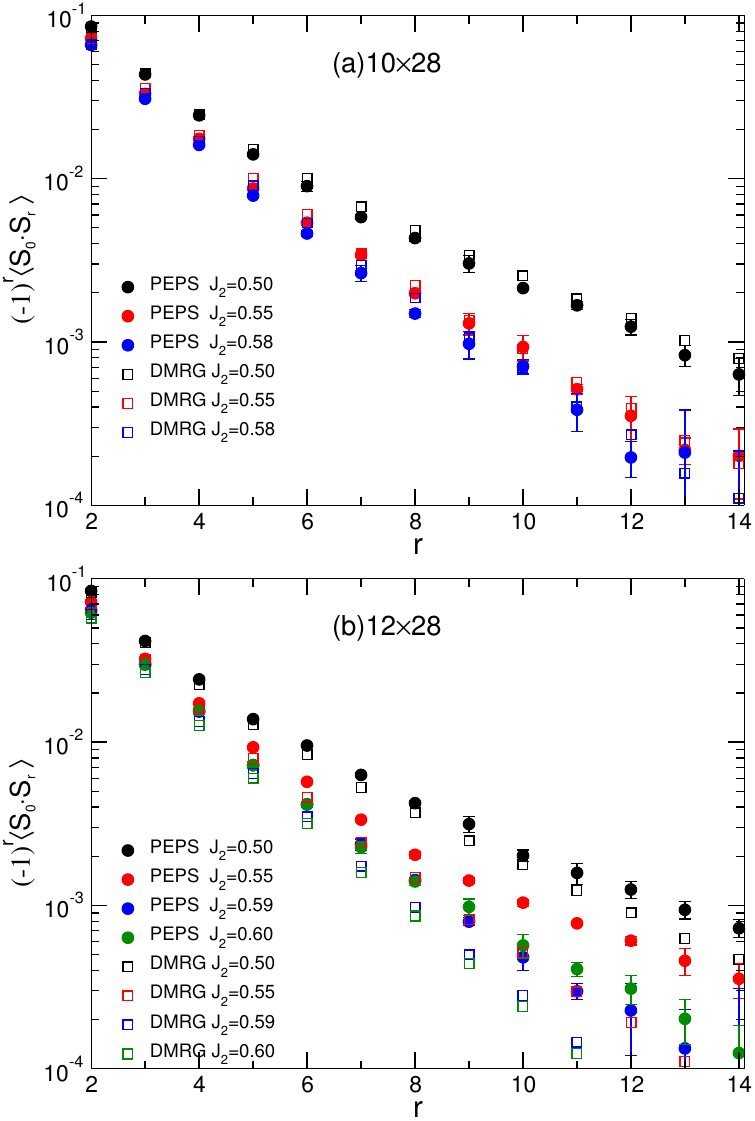}
 \caption{Comparison of the spin correlations vs distance obtained in DMRG and PEPS for different values of $J_2$ (semi-log plot).  $10\times28$ (a) and $12
\times28$ (b) strips have been considered. For $10\times28$, we use $M$=10000. For $12\times 28$,  we  push $M=14000$ at $J_2=0.55$, and $M=12000$ otherwise. PEPS results are obtained with $D=8$.}
 \label{fig:DMRGvsPEPS}
 \end{figure}

  \begin{figure}[htbp]
 \centering
 \includegraphics[width=3.4in]{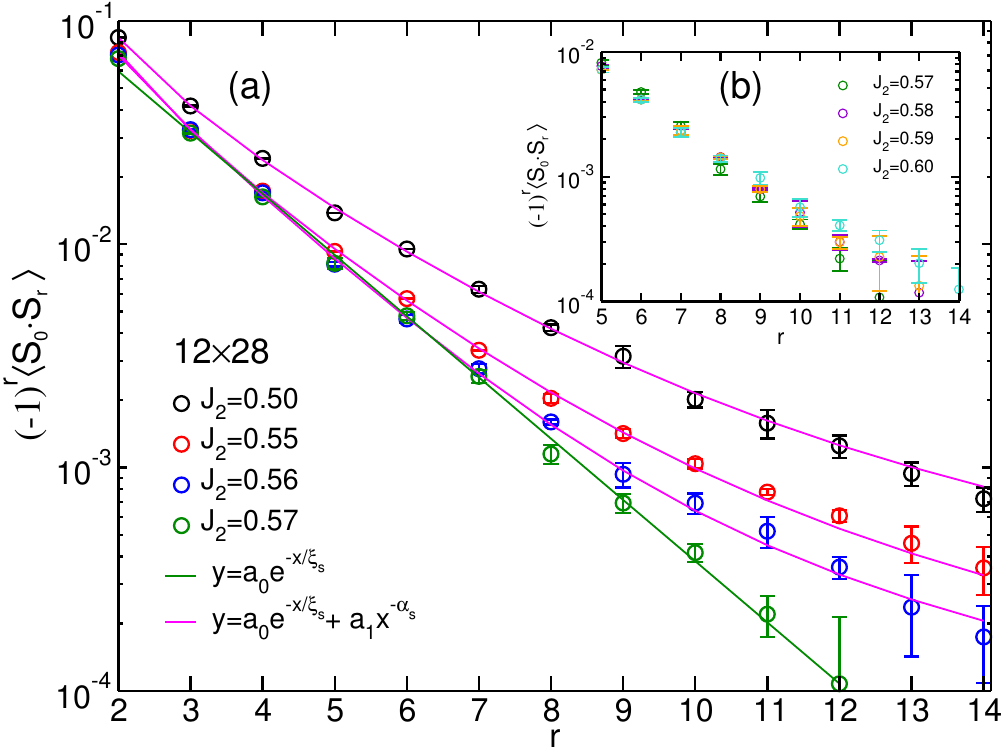}
 \caption{Spin correlation functions on a $12\times28$ strip computed using a $D=8$ PEPS at different values of $J_2$, $J_2\le 0.57$ (a) and $J_2\ge0.57$ (b) (semi-log plots). In (a) different functions are used to fit the spin correlations. The parameters from a two-component fit $y=a_0e^{-x/\xi_s}+a_1x^{-\alpha_s}$  used for $J_2=0.5$ and 0.55 are listed in Table~\ref{tab:twocomponentfit}. For $J_2=0.57$ the power-law component turns out to have negligible weight and the purely exponential fit $y=a_0e^{-x/\xi_s}$ gives a correlation length $\xi_s=1.59(2)$.  }
  \label{fig:Diffspin12x28}
 \end{figure}

  \begin{figure*}[htbp]
 \centering
 \includegraphics[width=6.8in]{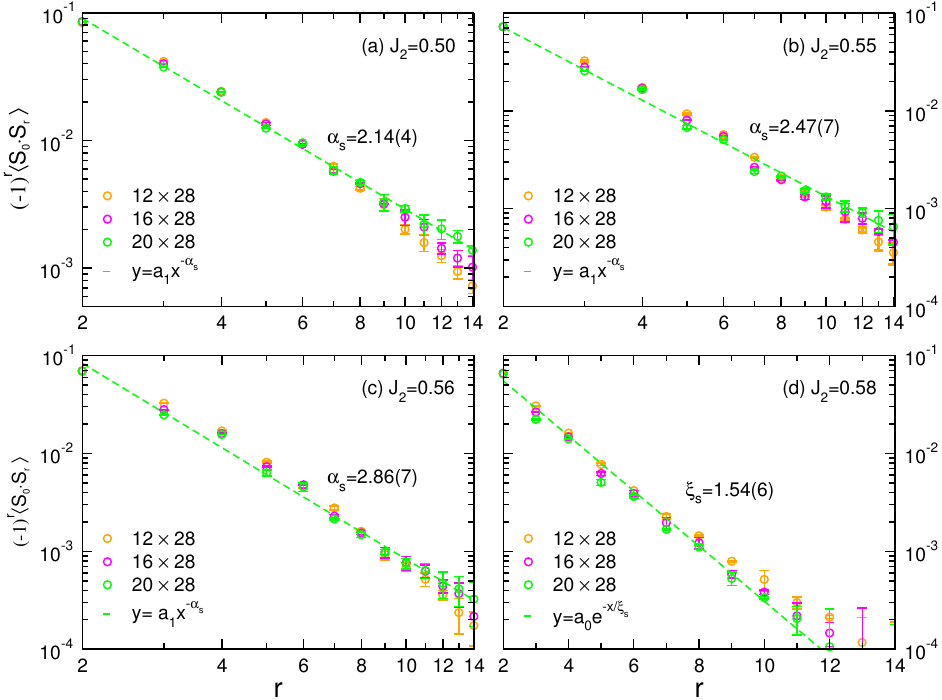}
 \caption{Log-log plot of the spin correlation functions on different system size at  (a) $J_2=0.5$, (b) $J_2=0.55$ and (c) $J_2=0.56$, and semi-log plot at (d) $J_2=0.58$. Different symbols correspond to different widths $L_y$ of the $L_y\times 28$ open strip. Black dashed lines denote the fitting curves of $20\times 28$ results with power law fit for (a,b,c) and exponential fit for (d). Corresponding fitting parameters $\alpha_s$ or $\xi_s$ are given in the figure. The parameters from a two-component fit $y=a_0e^{-x/\xi_s}+a_1x^{-\alpha_s}$ (solid lines) used for $J_2=0.5$ and $0.55$ (a,b) are listed in Table~\ref{tab:twocomponentfit}.  }
 \label{fig:AllSpinCorrDiffSize}
 \end{figure*}

Figure \ref{fig:AllSpinCorr} depicts the spin correlations versus distance on different systems at two highly frustrated points $J_2=0.5$ and $0.55$ on $10\times28$ and $12\times 28$ systems. DMRG  results are obtained with different numbers of SU(2) kept state, as large as $M=14000$ (equivalent to about 56000 U(1) states), and PEPS results are obtained with bond dimension $D$ from 6 to 10. The corresponding ground state energies for different $M$ or $D$ are listed with each legend showing that PEPS and DMRG  energies are very close to each other. 

For both DMRG and PEPS, by increasing the bond dimensions $M$ or $D$ one improves energies and correlations until convergence. To be more specific, we first focus on PEPS spin correlations. On $10\times 28$ and $12\times 28$ strips, with increasing $D$ from 6 to 10, spin correlations gradually increase but there are few differences between $D=8$ and $D=10$, which indicates $D=8$ already provides converged spin correlations. DMRG correlations also gradually increase with increasing $M$, and it needs larger $M$ to converge the long-distance correlations. Remarkably, as shown in Fig.~\ref{fig:AllSpinCorr}(a) and (b), once convergenced, PEPS and DMRG results on the $L_y=10$ strip are in excellent agreement. 
  
 However, we find that for $L_y$=12, some discrepancies between PEPS and DMRG correlations occur, even when considering the largest bond dimensions $M$ available in DMRG. At $J_2=0.5$ there are only small differences for long-distance correlations, but at $J_2=0.55$ differences occur already at short-distances while, at long-distance, DMRG correlations are definitely far away from the PEPS results, even when as many as $M=14000$ SU(2) states are kept, as seen in Fig.~\ref{fig:AllSpinCorr}(d).  Since the DMRG correlations have not totally converged, we attempt to extrapolate to $M\rightarrow \infty$ with the DMRG correlations at $M=10000$ and $M=14000$ for $J_2=0.55$ (or $M=12000$ for $J_2=0.5$) through a linear fit in $1/M$, as shown by the green symbols in Fig.~\ref{fig:AllSpinCorr}(c) and (d). For $J_2=0.5$ the extrapolated values agree very well with the converged PEPS results. For $J_2=0.55$ the extrapolated short-distance correlations also agree well with the PEPS results, while the extrapolated long-distance values are still significantly away, as a result of the inaccuracy of the DMRG extrapolation procedure.   
 Essentially, as a one-dimensional method, DMRG can not capture the correct  entanglement structure for large 2D systems, and its accuracy depends strongly on the width $L_y$ and the strength of the frustrating $J_2$ interaction (here $J_1$ is fixed). Increasing $L_y$ and $J_2$, one needs to keep more states to converge long-distance correlations as well as energies. As the extension of DMRG to higher dimensions, PEPS can capture the correct entanglement structure for 2D systems which satisfies the entanglement entropy's area law. Therefore, even though PEPS energies are slightly higher, $D=8$ can still produce  well-converged correlations at long distance. 
 
 To further investigate the influence of $L_y$ and $J_2$ on the accuracy of DMRG, we compare  PEPS spin correlations to DMRG results obtained with the largest available $M$, for different $J_2$. On $10\times28$ we can see PEPS results are in good agreement with DMRG results obtained with $M=10000$ for $J_2=0.5$, 0.55 and 0.58, as shown in Fig.~\ref{fig:DMRGvsPEPS}(a). When increasing  $L_y$ to 12, as shown in Fig.~\ref{fig:DMRGvsPEPS}(b), at $J_2=0.5$ DMRG and PEPS results correlations still compare well, but for $J_2=0.55$, 0.59 and 0.60, discrepancies occur due to lack of convergence of DMRG.  In fact, even for the largest available $M$, DMRG tends to underestimate long range spin correlations and, e.g. , gives a clear exponential decay in the suspected critical QSL at $J_2=0.55$ where, in contrast, well converged PEPS results are closer to the expected 2D power-law decay (see later for more results). This illustrates the powerful representation ability of PEPS for capturing long-range correlations in large 2D systems.  Thus, our results provide the first solid PEPS calculation beyond DMRG. 
 
\subsubsection{Power law decay behavior at long distance}

 From Fig.~\ref{fig:AllSpinCorr}(c) and (d), we note the spin correlations decay much more like in a power law, rather than an exponential form. In order to evidently establish their decay behavior, we plot the spin correlations on the $12\times 28$ open strip from $J_2=0.5$ to $0.57$ in Fig.~\ref{fig:Diffspin12x28}(a), and from $J_2=0.57$ to $0.60$  in Fig.~\ref{fig:Diffspin12x28}(b). When increasing $J_2$ from 0.5 to 0.57, the correlations gradually get smaller. Interestingly, under further increasing $J_2$ to 0.60, the correlations  seem to get back a little larger. 
 To analyze the correlation behavior in details, we fit the correlations with two different fitting functions, as shown in Fig.~\ref{fig:Diffspin12x28}(a). Comparing the two kinds of fitting functions, we can clearly see that (i) at $J_2=0.57$ spin correlations decay exponentially, while (ii) for $J_2=0.5$, 0.55 and 0.56 the spin correlations decay with a long tail indicating a power law form.

  \begin{table}[htbp]
   \centering
 \caption {Fitting parameters  of the spin-spin correlations on $L_y\times 28$ open strips, using the function $y=a_0e^{-x/\xi}+a_1x^{-\alpha}$. }
	\begin{tabular*}{\hsize}{@{}@{\extracolsep{\fill}}llllll@{}}
		\hline\hline   
	 $J_2$ & $ L_y$ &  $a_0$ & $a_1$    & $\xi_s$ &$\alpha_s$  \\ \hline
 		0.50 & 12 & 0.095(9)  & 0.22(2)   & 1.99(9)  & 2.16(9)  \\
 				0.50 & 16 & 0.064(13)  & 0.27(3)   & 2.07(28)  & 2.16(13)  \\
 						0.50 & 20 & 0.032(18)  & 0.31(4)   & 2.29(98)  & 2.12(21)  \\
 		0.55 & 12 &0.107(6)  &0.22(1)   & 1.69(6)  & 2.50(8)  \\
 			0.55 & 16 &0.043(29)  &0.32(7)   & 1.74(73)  & 2.45(35)  \\
 				0.55 & 20 &0.004(4)  &0.39(2)   & 1.89(87)  & 2.45(9)  \\		
 		\hline\hline
	\end{tabular*}
\label{tab:twocomponentfit}	
\end{table}

 To provide more evidences on the change of behavior of the long distance spin correlations with $J_2$, we also consider larger systems up to $20\times28$. In Fig.~\ref{fig:AllSpinCorrDiffSize}, we present how the spin correlations vary with increasing system width $L_y$.  We can see, when increasing $L_y$, two different trends of the long-distance spin correlations~: at $J_2=0.5$, 0.55 and 0.56 the latter increase, while at $J_2=0.58$ they decrease. This consolidates our claim of a power law decay behavior for $J_2=0.5$, 0.55 and 0.56, and an exponential decay for $J_2=0.58$. The power exponents for $J_2=0.5$ and 0.55 obtained from the widest $20\times 28$ strip are $\alpha_s=2.14(4)$ and $2.47(7)$ respectively, in good agreement with the previous results obtained on the $12\times 28$ strip through a two-component fit. The correlation length  at $J_2=0.58$ for $20\times 28$ is $\xi_s=1.54(6)$, well consistent with the results from $12\times 28$ at $J_2=0.57$ \footnote{the estimation of $\xi_s$ at $J_2=0.57$ is subject to finite-$L_y$ effects and that the thermodynamic value of $\xi_s$ is expected to be significantly larger than
the value extracted from Fig.~\ref{fig:Diffspin12x28}(a) for $L_y=12$, as is seen from Fig.~\ref{fig:AllSpinCorrJ2}(f).}. Such results on wide strips confirm and strengthen the previous findings of a gapless QSL and a (gapped) VBS state in the nonmagnetic region, and a phase  boundary between the latter at $J_{c2}=0.56$.

  \begin{figure}[htbp]
 \centering
 \includegraphics[width=3.4in]{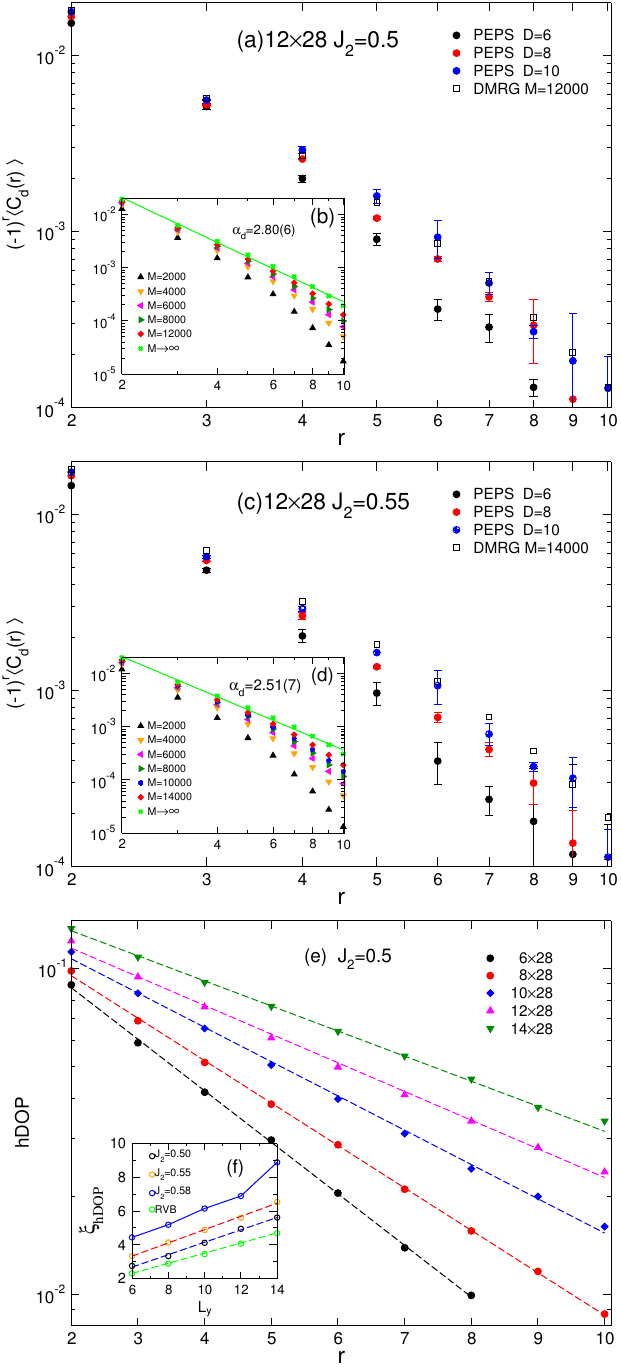}
 \caption{Log-log plot of dimer correlation functions for different system sizes at  (a) $J_2$=0.5 and (c) $J_2$=0.55.  The insets (b) and (d) show the DMRG results with different numbers of kept SU(2) states, and the green lines denote the extrapolated DMRG results in the $M\rightarrow \infty$ limit (using a linear fit in $1/M$ and the two largest $M$ values) which are fitted according to a power law in $1/r$. The corresponding exponents $\alpha_d$ are indicated in the insets.}
 \label{fig:AllDimerCorr}
 \end{figure}

\subsection{Dimer-dimer correlation functions}
To further study the physical properties of the QSL, we also measure the {\it connected} dimer-dimer correlations on the central line along the $x$-direction, which is defined as 
\begin{equation}
C_d(r)=\langle B^x_0B^x_r \rangle-\langle B^x_0\rangle\langle B^x_r \rangle \, .
\end{equation}
The comparison between DMRG and PEPS at $J_2=0.5$ and $J_2=0.55$  can be found in Fig.~\ref{fig:AllDimerCorr}. We can see that with increasing $D$ or $M$, PEPS and DMRG dimer-dimer correlations will gradually converge, and results of PEPS with $D=10$  are  consistent with those of DMRG with the largest available $M$.
 However, dimer correlations are significantly  smaller than spin correlations and are slower to converge in both DMRG and PEPS calculations. In order to estimate the dimer correlations more precisely, we can also extrapolate the dimer correlation values with $M\rightarrow \infty$ from DMRG results, as seen in the insets of  Fig.~\ref{fig:AllDimerCorr} (b) and (d). From the extrapolated DMRG results, we conclude that the dimer-dimer correlations very likely decay as a power law, and exponents $\alpha_d=2.80(6)$ and 2.51(7) have been estimated for $J_2=0.5$ and 0.55, respectively.

\begin{figure}[htbp]
 \centering
 \includegraphics[width=3.4in]{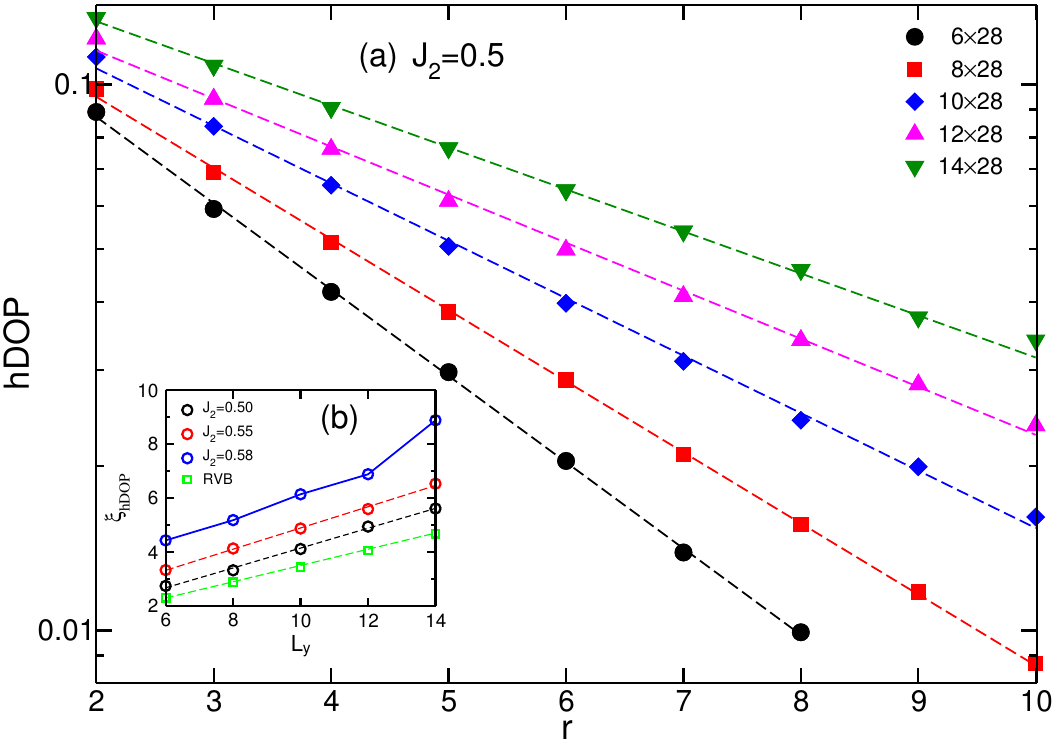}
 \caption{(a) hDOP obtained on different $L_y\times 28$ open strips at $J_2=0.5$, plotted as a function of the distance from the reference site along the central line $y=L_y/2$.  The corresponding decay lengths for several width $L_y$ are extracted from exponential fits (shown as straight lines). (b) The decay length of the hDOP at $J_2=0.5$ and $0.55$ grows linearly with the system width, like for the RVB state. The slopes for RVB, $J_2=0.5$ and $0.55$ are 0.301(4), 0.37(1) and 0.39(1), respectively. At $J_2=0.58$ a (small) sign of superlinearity is seen. }
 \label{fig:BondHdop}
 \end{figure}

In addition, we also investigate the system size dependence of the local DOP and of its characteristic decay length.  The horizontal DOP (hDOP) is defined as the difference $\Delta B$ between nearby strong and weak horizontal bond energies
\begin{equation}
\Delta B (r) =|B^x_r - B^x_{r+e_x}|   \, .
\end{equation}
 The hDOP decays exponentially from the left system boundary and the corresponding decay length $\xi_{\rm hDOP}$ can be extracted, as shown in Fig.~\ref{fig:BondHdop}(a). We find that, at $J_2=0.5$ and 0.55, the decay length grows linearly with increasing system size, exhibiting the same behavior as that of the RVB state. This is consistent with a power law decay of the dimer-dimer correlation functions in the QSL phase. In contrast, at $J_2=0.58$, the growth seems faster, may be superlinear, consistent with the appearance of VBS order in the thermodynamic limit.

\section{A potential field theory description for the quantum spin liquid phase}
To understand the numerical results qualitatively, we start from the well known $CP^1$ model description for DQCP between AFM and VBS phases:
\begin{eqnarray}
\mathcal{L}_{\text{DQCP}}&=&\frac{1}{g}\sum_{a=1}^2 |(\partial_\mu-i a_\mu)z_a|^2+m^2|z|^2+u(|z|^2)^2\nonumber\\ &+&\kappa (\epsilon_{\mu\nu\lambda} \partial_\nu a_\lambda)^2,  \label{DQCP}
\end{eqnarray}
where $z_a$ is the $CP^1$ spinon field and $a_\mu$ is the emergent gauge field. 

Our numerical results indicate that the QSL state is closely related to the DQCP and it can be regarded as a natural expansion of the DQCP into a stable quantum phase. Then the natural question would be: is there any instability for the usual DQCP scenario such that a stable QSL phase might emerge around it? Here, we conjecture that a topological theta term, or the Hopf term: 
\begin{eqnarray}
\mathcal{L}_\theta=\frac{\theta}{4\pi^2}\epsilon_{\mu\nu\lambda} a_\mu \partial_\nu a_\lambda 
\end{eqnarray}
with $\theta=\pi$ might do the job. Although in the limit with $J_2=0$, it was proven that there is no topological theta term contribution in the usual 2D AFM Heisenberg model~\cite{Haldane}, it is still not clear whether such a term can emerge or not in the presence of bigger $J_2$.

\begin{figure}[htbp]
 \centering
 \includegraphics[width=3.4in]{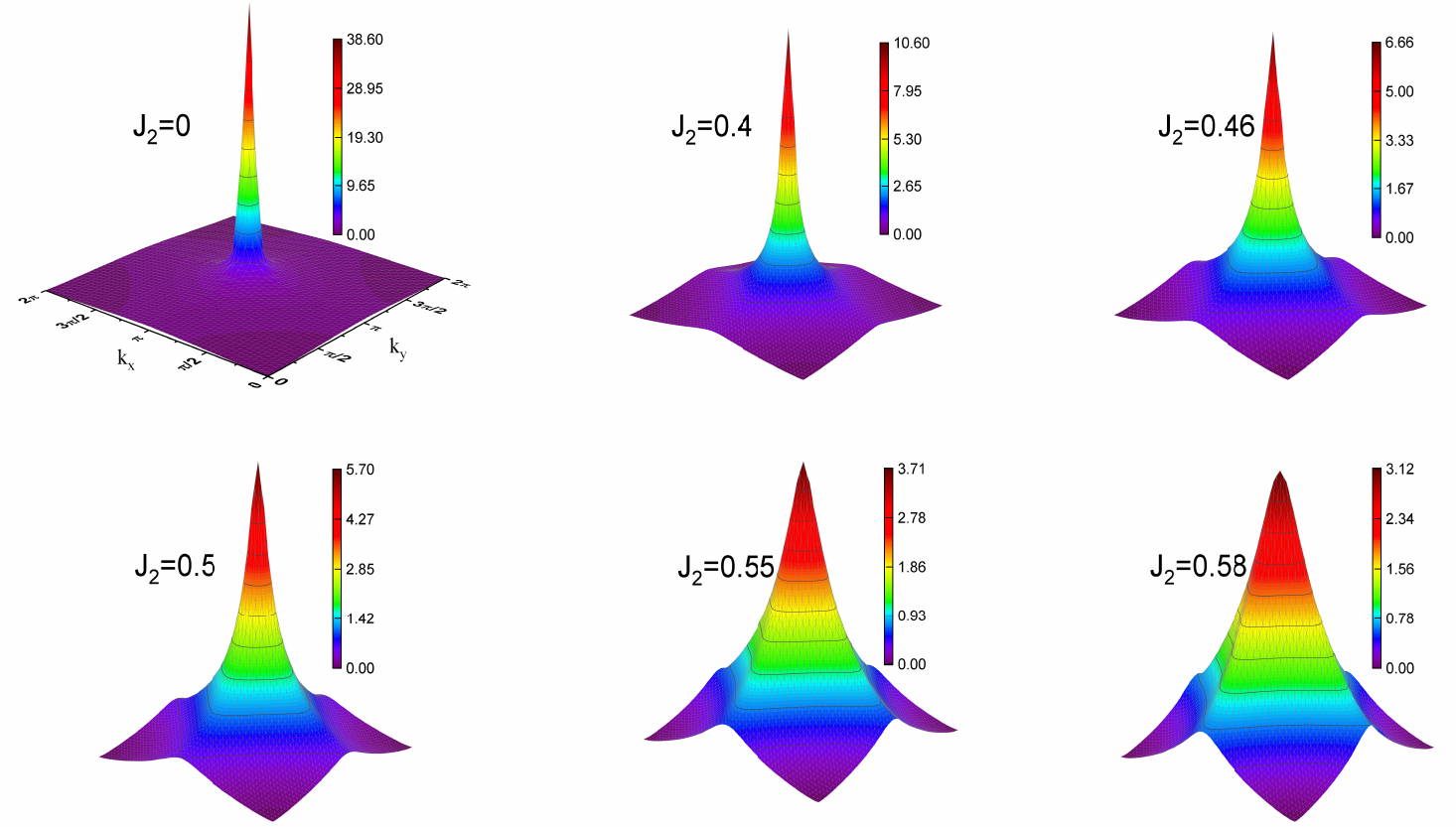}
 \caption{The $\bf k$-space distribution of spin structure factor $S({\bf k})=L^2m^2_s({\bf k})$ at different $J_2$ on $L\times L$ systems with $L = 20$. Each case shares the same  $k_x$ and $k_y$ axes with the $J_2=0$ case, with a peak at ${\bf k}=(\pi,\pi)$. }
 \label{fig:structurefactor}
 \end{figure}

According to Polyakov's early work~\cite{Polyakov}, in the presence of $\mathcal{L}_\theta$, the total action $\mathcal{L}_{\text{total}}=\mathcal{L}_\text{DQCP}+\mathcal{L}_\theta$ is equivalent to four Dirac fermions with short-range interactions. In terms of physical picture, the topological theta term $\mathcal{L}_\theta$ leads to the statistical transmutation of spinon excitation in the $CP^1$ model. Thus, a power law decay of the spin-spin correlations will be expected at long distance. Due to the square lattice symmetry, the four Dirac spinons will naturally locate at momenta $(\pm \pi/2, \pm \pi/2)$.
In Fig.\ref{fig:structurefactor}, we compute the spin structure factor for different $J_2$ and we find that subpeaks indeed emerge at momentum point $(0, \pi)$ and $(\pi, 0)$ in the QSL regime. We argue that these subpeaks in the spin structure factor might be naturally contributed by those Dirac spinons located at momenta $(\pm \pi/2, \pm \pi/2)$.

On the other hand, since $\mathcal{L}_\theta$ will not change the short range physics of the $CP^1$ model, the short distance behavior of correlation functions can still be captured by the usual $CP^1$ model without $\mathcal{L}_\theta$ and that's why the observed short range spin-spin and dimer-dimer correlations are qualitatively similar to those in the RVB state. (With a two component fitting, we can always find an exponetial decay part for spin-spin correlations for small system size.) In previous works~\cite{WangRVB,YangJ1J2,poilblanc2017}, it has been shown that the RVB state indeed has a very good variational energy around the maximally frustrated region with $J_2=0.5$. Thus, we argue that the RVB state can be regarded as a meta-stable state and it naturally serves as the "parent state" for gapless QSL state with power law decay of spin-spin correlation functions.

It is well known that in the DQCP scenario, the bigger anomalous dimension $\eta_s$ can be rationalized by the thinking that the Néel order parameter field decays into spinons. Remarkably, we find that $\eta_s$ is intrinsically close (comparing systems of similar sizes) to the $J$-$Q$ model value around the AFM-QSL transitions. While for the QSL-VBS transition, $\eta_d$ obtained in our numerical calculation also well agrees with the $J$-$Q$ model(again, comparing with similar system size). These two features strongly indicate that the observed QSL might naturally develop from an underlying DQCP.
Moreover, we find that $\eta_s$ for the QSL-VBS transition is much bigger than the the $J$-$Q$ model and is actually very close to the infinite-$N$ limit of $CP^{N-1}$ model with $\eta_s=1$ for QSL-VBS phase transitions. We conjecture that the statistical interaction induced by $\mathcal{L}_\theta$ might strongly suppress the quantum fluctuation of $U(1)$ gauge field at QSL-VBS phase transition(confinement transition) and lead to almost "free" fermionic spinon excitations.
We will use the large-$N$ expansion to estimate $\eta_s$ in the presence of $\mathcal{L}_\theta$ term elsewhere. Finally, we stress that the observed correlation exponents $\nu \sim 0.99$  for both AFM-QSL and QSL-VBS transitions might imply the deep relationship between Higgs and confinement transitions of $U(1)$ gauge field. Since $\nu$ is much bigger than those observed in the usual DQCP scenario (comparing systems of similar sizes), e.g., $J$-$Q$ model, we believe that the observed QSL can not be a finite size effect and both transitions must belong to new universality classes which are never observed in other models.  
Of course, doping the square lattice $J_1$-$J_2$ model might provide us smoking gun evidence for the emergence of topological theta term and statistic transmutation for spinons, and a d-wave superconductivity could naturally arise. We will also leave this interesting open problem for our future study. 

\bigskip
\section{discussion and conclusion}
In summary, applying the state-of-the-art PEPS method to the frustrated $J_1$-$J_2$ model up to $24\times 24$ open systems, we compute the ground state energies, order parameters and correlation functions with unprecedented precision. Through the analysis of finite-size scaling, our results explicitly show that  in the nonmagnetic region $0.45\lesssim J_2 \leq 0.61$, there exists a gapless QSL phase for $0.45\lesssim J_2 \lesssim 0.56$  and a  weak VBS phase for $0.56\lesssim J_2 \leq 0.61$.  This phase diagram is further supported by the behavior of the spin correlations on wide open strips, which decay as a power law at $J_2=0.5$ and 0.55  and exponentially at $J_2=0.58$. Besides, the dimer correlations also decay as a power law in the gapless QSL.  Furthermore, we also fit the critical exponents at the AFM-QSL and QSL-VBS transitions, which strongly indicates new types of universality class for these two transitions. Finally, we provide a potential quantum field theory framework to understand the physical nature of gapless QSL. We would like to stress that the proposed gapless QSL also gives a concrete example beyond the usual PSG framework. 

We stress that our results are among the most reliable on this topic. Besides the excellent accuracy, the very high quanlity of our results is further guaranted by a series of cross checks. These cross checks include: (1) To locate the vanishing point of AFM order, we use the method of finite size scaling analysis of AFM order parameters and the method of the crossing point of a dimensionless quantity $\xi_m/L$; (2) To detect the possibility of VBS phase, we use two kinds of VBS order parameters, the one induced by boundaries and the one defined by bond-bond correlations; (3) To locate the VBS-stripe phase transition point, we use stripe order parameter and ground state energy as indepedent approaches. Even for the ground state energy approach, we still use the energy peak position and bulk energy two methods; (4) To determine the decay behaviour of spin-spin correlations, we compute the variation of correlations with system size increasing and we also check how correlations evolve with J2 increasing; (5) To determine the decay behaviour of dimer-dimer correlations, we compute connetced dimer-dimer correlation functions, and we also use the hDOP quantities for further check; (6) We also use different bond dimensions to check the convergence of our results. PEPS and DMRG results, and decay behaviour of correlations and the phase diagrams, respectively, also consist of double check; (7) In the whole process, we also use a well-studied gapless RVB spin liquid state as a reference.

Our high-quality results enable us to put together in a consistent way incomplete and/or seemingly conflicting findings from previous studies. Former vQMC~\cite{hu2013} and recent finite PEPS \cite{liu2018} studies suggest a gapless QSL but these studies mainly focused on the region $0.4\leq J_2 \leq 0.55$, and the nature of the ground state in the region $0.55 < J_2 \leq 0.62$  is not carefully studied. An infinite-PEPS (iPEPS) study supports a VBS phase for $0.53\lesssim J_2 \lesssim 0.61$, but the AFM order vanishes at about $J_2=0.53$~\cite{CVB4} which is probably caused by not large enough bond dimension $D$~: iPEPS typically tends to overestimate magnetic correlations and a finite $D$ may induce a spurious finite magnetic order~\cite{liao2017}.  However, new finite correlation length scaling in iPEPS provides very good results~\cite{corboz2018,rader2018}, giving a critical value around $J_{c1}\simeq 0.46(1)$~\cite{juraj2020}, in agreement with our findings. It is also very challenging for iPEPS  to obtain a gapless QSL~\cite{poilblanc2017}, often requiring a very large $D$ to be reached~\cite{liao2017}.

 In the SU(2)-DMRG study of Ref.~\cite{gong2014}, a VBS phase is proposed for $0.5 \lesssim J_2 \lesssim 0.61$, associated with a near critical region $0.44 \lesssim J_2 \lesssim 0.5$. Later, phase diagrams including both spin liquid and VBS phases are suggested in a many-variable vQMC study~\cite{morita2015}, which reports a continuous transition between the spin liquid and the VBS at $J_{c2}\simeq 0.5$. Such a  phase diagram is also obtained by an indirect approach on the basis of energy level crossing analysis with the most recent DMRG calculations~\cite{wang2018} and another vQMC cacluations~\cite{ferrari2020}, while no clear sign of dimer order is visible in the correlation functions~\cite{ferrari2020}.  Our accurate tensor network results up to $24\times 24$ systems evidently establish  that both gapless QSL and VBS phase do exist in the nonmagnetic region, supported by the decay behavior of  spin-spin and dimer-dimer correlations. Furthermore, we have been able to obtain the critical exponents of the AFM-QSL and QSL-VBS transitions for understanding the nature of the phases.

Very recently, a similar phase diagram was also reached using neural network wave functions~\cite{nomura2021}, i.e., the RBM+PP method, although providing slightly different values of the critical points. In particular, 
the QSL-VBS transition point is estimated to be $J_{c2}=0.54$ -- instead of $J_{c2}=0.56$ here -- which seems consistent with another iPEPS result suggesting a very weak VBS (as revealed by a very long dimer correlation length) at $J_2=0.55$~\cite{poilblanc2019}. However, without considering a relatively high estimated RBM+PP energy in 2D limit $E_{\rm RBM+PP}\simeq-0.4846$, the RBM+PP energy on large sytems like $16\times 16$ torus $E=-0.4852$ is already higher than our 2D limit energy $E_{\rm PEPS}\simeq -0.4857$, which actually indicates the large size results of RBM+PP are not so accurate as their small size results. Our finite PEPS results, including finite size scaling of order parameters, and the behavior of spin and dimer correlations, all support that the ground state at $J_2=0.55$ is critical, consistently with the earlier vQMC results~\cite{hu2013}. We note in Ref.~\cite{nomura2021} the obtained spin and dimer critical exponents  $z+\eta_{s1}\approx 1.41$ and $z+\eta_{d1}\approx 1.75$ at the AFM-QSL transition point, and $z+\eta_{s2}\approx 1.60$ and $z+\eta_{d2}\approx 1.44$ at the QSL-VBS transition point,  are roughly consistent with our results  $z+\eta_{s1}\approx1.38$ and $z+\eta_{d1}\approx1.72$ at the AFM-QSL transition point, $z+\eta_{s2}\approx1.96$ and $z+\eta_{d2}\approx1.26$ at the QSL-VBS transition point. In Ref.~\cite{nomura2021}, two different correlation length exponents: $\nu=1.21(5)$ at the AFM-QSL and $\nu=0.67(2)$ at the QSL-VBS transition point~\cite{nomura2021}, are obtained by scaling order parameters, but our results find a single correlation length critical exponent  $\nu=0.99(6)$ at the two transition points. The disagreement might be caused by the potential  accuracy problem on large sizes, as well as the fitting way to extract $\nu$.  In Ref.~\cite{nomura2021}, the exponents $z+\eta$ and $\nu$ are fitted by simultaneously adjusting these exponents through scaling the order parameters. In our work, we use the method suggested in Ref.~\cite{FSS1} to extract the exponent $\nu$ first by scaling the correlation length itself, which can avoid simultaneously adjusting other critical exponents.  Then a single $\nu=0.99(6)$ is further supported by scaling order parameters. It would be interesting to check the exponent $\nu$ by directly extracting from the correlation length itself in Ref.~\cite{nomura2021}. We would like to point out that from several other frustrated models that also posses AFM-QSL and QSL-VBS transitions, we always observes  a single correlation length exponent $\nu\approx 1.0$, which will be published elsewhere~\cite{j1j2j3}. 

Lastly, we note the QSL phase is critical and our computations have been done at a fixed bond dimension $D=8$ and, hence,  let us briefly discuss the effect of a finite $D$. Like DMRG, PEPS is belived to work well on the systems whose ground states satisfy the entanglement entropy's area law by using a (not very large) finite $D$. However, for those systems that do not satisfy the area law including gapless systems, the language of PEPS can also apply well  just like what DMRG has ever done. The basic idea is to improve the bond dimension as large as possible and explore reasonable extrapolation techniques. Along this line, the iPEPS method has moved forward a lot by designing efficient algorithms~\cite{jiang2008,roman2009,corboz2011,xie2014,phien2015,xie2017,fishman2018,liao2019} and developing new theories for extrapolations~\cite{corboz2018,rader2018}. Besides the popular iPEPS method, the finite PEPS approach here we adopt provides an alternative way. In this context, we aim to accurately simulate the systems as large as possible within the realm of our capability and then use  available results for finite size scaling to measure the thermodymic limit properties.  Since finite systems would always acquire a gap, we believe there must be a function $D_{\rm min}(L)$ that fixes the minimum $D$ to get D-converged results for a given (linear) system size $L$. We believe $D=8$ here is larger than the necessary bond dimension $D_{\rm min}(L)$ for the largest $L=28$ considered here, so that results are well converged. However, considering larger systems (for even higher accuracy) may require larger $D$ to be converged.

Methodologically, our PEPS calculations provide the first example beyond DMRG of high precision investigation of the  $J_1$-$J_2$ model through one-to-one direct benchmark for small system sizes.  Both DMRG and PEPS are theoretically unbiased approaches to deal with finite size frustrated models. However, DMRG is strongly limited to 1D and quasi-1D systems. PEPS is designed for two and higher dimensions, but its power has not been fully demonstrated in numerical simulation until now. The DMRG and PEPS comparisons explicitly show how DMRG gradually fails in 2D, and meanwhile numerically demonstrate the powerful representation ability of PEPS for precisely capturing long-range physics for such highly frustrated 2D systems.  It also fills the gap between DMRG and PEPS calculations which always appear separately in previous studies, and thus provides an enlightening approach to understand and check the discrepancies between existing DMRG and 2D tensor network  results that are contradictory for other models, which could be crucial to clarify the true nature of a series of controversial quantum many-body problems. The PEPS application here  is  an excellent prototype on solving long-standing 2D strongly correlated quantum many-body problems by tensor network methods, which we believe will have profound impact on the development of quantum many-body computations and theories. Experimentally, real materials for realizing $J_1$-$J_2$ model have been explored. A series of compounds with dominant $J_1$ and almost negligible $J_2$ have been investigated~\cite{ExpNeel1,ExpNeel2,ExpNeel3,ExpNeel4,ExpNeel5,ExpNeel6,ExpNeel7}, including the high-Tc superconductivity parent compound  $\rm La_2CuO_4$. The materials with weak $J_1$ and strong $J_2$ have also been found such as $\rm Sr_2CuWO_6$, $\rm Li_2COMO_4$ ($\rm M=Si,Ge$) and $\rm AMoOPO_4Cl$ ($\rm A= K, Rb$) in which collinear AFM order is observed~\cite{ExpStripe1,ExpStripe2,ExpStripe3,ExpStripe4,ExpStripe5,ExpStripe6}. But the experimental realization of  a $J_1$-$J_2$ model with the appropriate $J_2/J_1$ ratio for the highly frustrated region is still very scarce. We hope our theoretical work can attract more attention to further stimulate experimental development on this subject.

\section{acknowledgment}
We thank helpful discussions with Federico Becca, Ignacio Cirac, Juraj Hasik, Hai-Jun Liao, Yang Qi, Norbert Schuch, Dong-Ning Sheng,  Frank Verstraete, Qing-Rui Wang, Xiao-Gang Wen, Fan Yang, and Shuo Yang. We also thank  inspiring discussion with Anders Sandvik, and thank Yusuke Normura for helpful discussion and providing raw data. This work is supported by the NSFC/RGC Joint Research Scheme No. N-CUHK427/18, the ANR/RGC Joint Research Scheme No. A-CUHK402/18 from the Hong Kong's Research Grants Council and the TNSTRONG ANR-16-CE30-0025, TNTOP ANR-18-CE30-0026-01 grants awarded from the French Research Council. Shou-Shu Gong is supported by National Natural Science Foundation of China grants 11874078, 11834014, and the Fundamental Research Funds for the Central Universities. Wei-Qiang Chen is supported by National Key Research and Development Program of China (Grant No. 2016YFA0300300), NSFC (Grants No. 11861161001), the Science, Technology and Innovation Commission of Shenzhen Municipality (No. ZDSYS20190902092905285), and Center for Computational Science and Engineering at Southern University of Science and Technology..

\appendix
 \section{Energy convergence}
 
  \begin{figure}[htbp]
 \centering
 \includegraphics[width=3.4in]{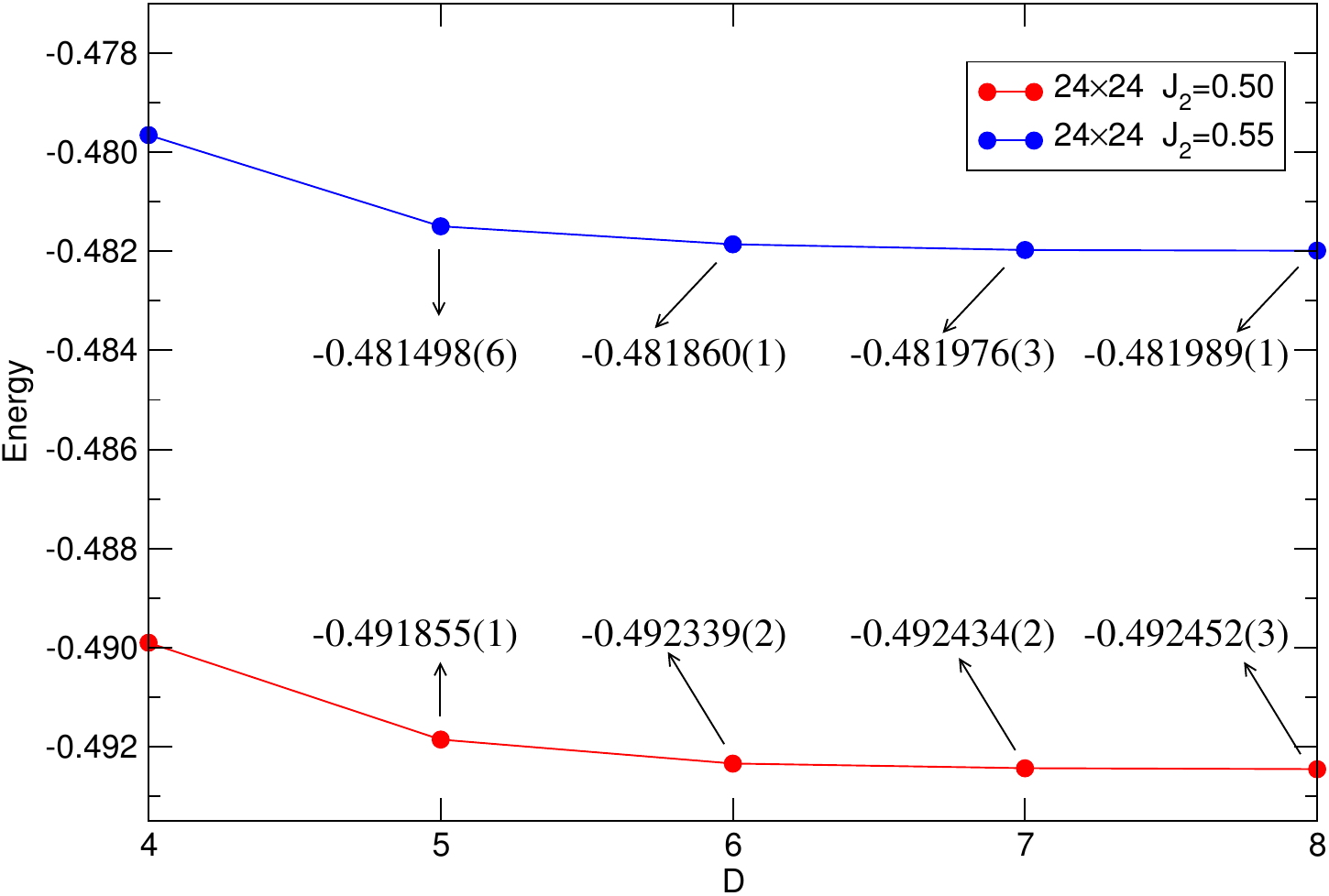}
 \caption{Energy convergence with respect to PEPS bond dimension $D$ on $24\times 24$ at $J_2=0.5$ and 0.55. Results using $D=5-8$  are listed.}
  \label{fig:energy}
 \end{figure}
 
We show the energy convergence with respect to the bond dimension $D$ on the used largest size $24\times 24$. We choose two typical highly frustrated points $J_2=0.5$ and 0.55 as examples, and compare the energies with $D=4-8$. From Fig.~\ref{fig:energy}, we can see when $D$ increases from 4 to 6, the energy decrease is visible. Further increasing $D$ the energy improvement is very small. For example,  at $J_2=0.55$, the energies with $D=7$ and $D=8$ respectively are $-0.481976(3)$ and $-0.481989(1)$.  This indicates $D=8$ can well converge the results. 

 \section{estimated 2D limit energy at $J_2=0.5$}
 \label{app:energyJ2_0.5}
\begin{figure}[htbp]
 \centering
 \includegraphics[width=3.4in]{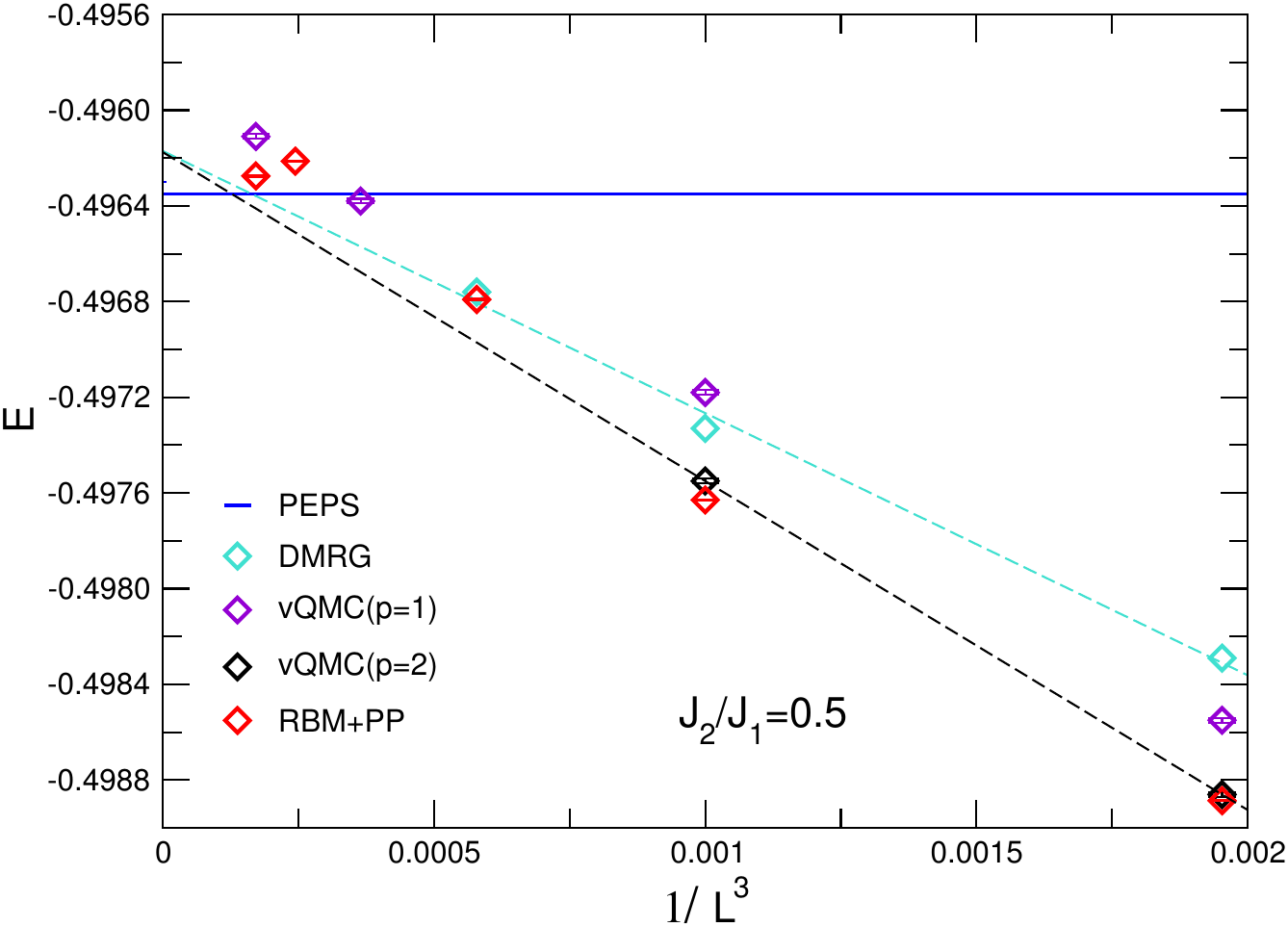}
 \caption{The ground state energy variation with respect to $1/L^3$. PEPS 2D limit energy (solid blue line) is shown for comparison. DMRG energies with $L=8-12$  taken from Ref.~\cite{gong2014} use cylindrical boundary condition. vQMC p=1 results with $L=8-18$ and p=2 results with $L=8-10$ taken from Ref.~\cite{hu2013}, and RBM+PP results with $L=8-18$ taken from Ref.~\cite{nomura2021}, are based on periodic boundary conditions. Dashed lines denote  linear fits versus $1/L^3$ for DMRG (cyan) and vQMC p=2 energies (black). }
  \label{fig:energyJ2_0.5}
 \end{figure}

We compare the estimated 2D limit energy at $J_2=0.5$ from different methods. The 2D limit energy from finite PEPS  is $E_{\rm PEPS}=-0.49635(5)$ by finite-size scaling up to $24\times 24$ sites~\cite{liu2019}. From Fig.~\ref{fig:energyJ2_0.5}, we can see the energies from cylindrical and periodic boundary conditions are roughly consistent with a  $1/L^3$ leading scaling. With a linear fit of $1/L^3$, the DMRG using $L=8-12$ and vQMC using $L=8-10$ produce almost the same extrapolated energy about $-0.4962$, slightly higher than our finite PEPS 2D limit energy  $-0.49635(5)$. The RBM+PP energy on large sizes $L=16$ and $L=18$ are already higher than $E_{\rm PEPS}=-0.49635(5)$, indicating a higher 2D limit energy than $E_{\rm PEPS}$ if the energy increases monotonically with $L$ increasing. Note that $L=16$ energy is slightly higher than that of $L=18$ might be caused by different choices of sublattice structure, the former using $4\times 4$ sublattice and the latter using $6\times 6$ sublattice~\cite{nomura2021}.  In anyway, this demonstrates the finite PEPS results indeed have a very good accuracy.

\section{collinear AFM  phase transition point}
\label{app:stripe}
\begin{figure}[htbp]
 \centering
 \includegraphics[width=3.4in]{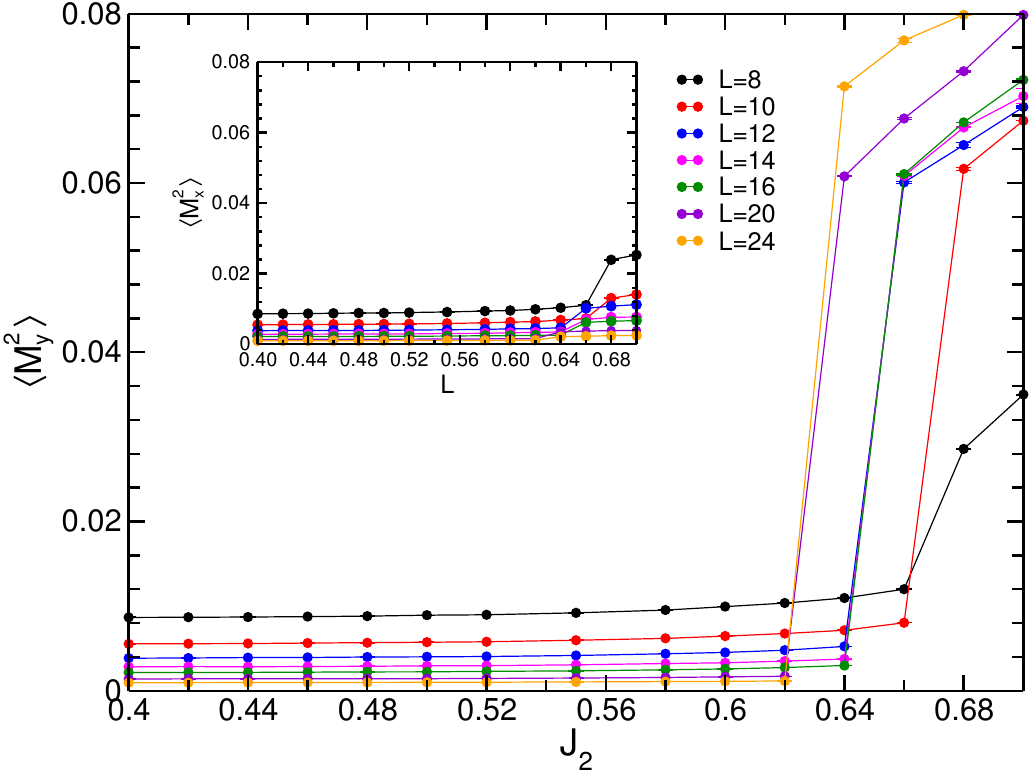}
 \caption{The collinear AFM order parameter $M^2_x=m^2_s({\pi,0})$ and $M^2_y=m^2_s({0,\pi})$ for different $J_2$ on $L\times L$ lattices up to $L=$24 and a PEPS with $D=8$. }
 \label{fig:stripeorder}
 \end{figure}

\begin{figure}[htbp]
 \centering
 \includegraphics[width=3.4in]{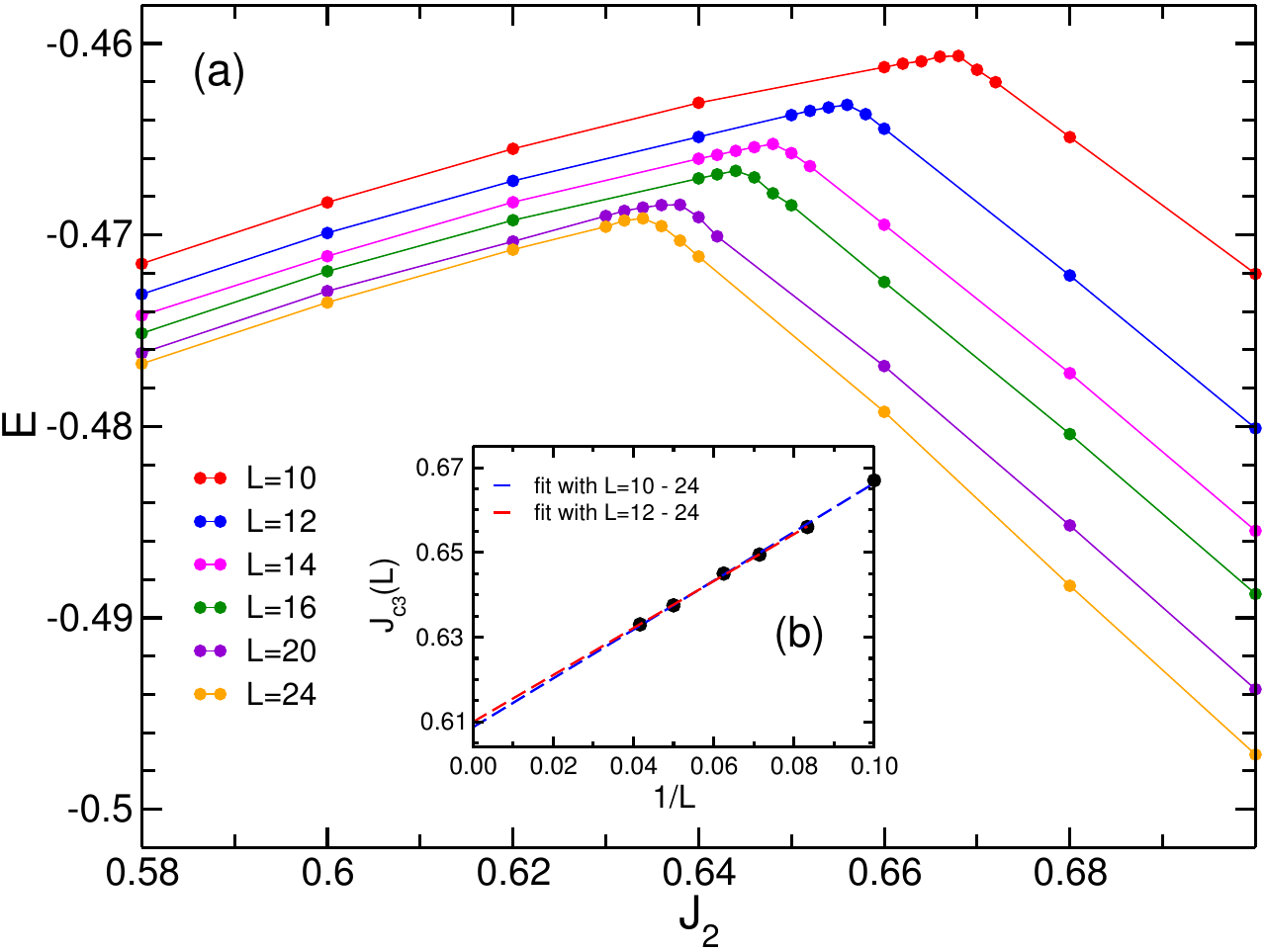}
 \caption{(a) The $J_2$ dependence of ground state energies obtained with a $D=8$ PEPS on an $L\times L$ open system, $L=10-24$. For each size, the phase transition point $J_{c2}(L)$ is corresponding to the peak position. (b) The estimation of the transition point  in the 2D limit obtained by finite-size scaling of $J_{c3}(L)$ versus $1/L$ through linear fits. The blue line denotes the fit for $L=10-24$ with an extrapolated value 0.609(1), and the red one for $L=12-24$ with an extrapolated value $J_{c3}=0.6100(3)$.}
 \label{fig:transitionPoint}
 \end{figure}

Figure~\ref{fig:stripeorder} depicts the collinear AFM order $M^2_x=m^2_s({\pi,0})$ and $M^2_y=m^2_s({0,\pi})$  in the region $0.4\leq J_2\leq 0.70$. We can see for each size $L\times L$, when $J_2$ increases and is larger than some value $J_{c3}(L)$  the system will experience a first-order phase transition and then goes into a collinear AFM phase, characterized by a sudden appearance of the collinear order parameter and explicit symmetry breaking  with inequivalent $x$ and $y$ directions. Note that the  transition point  $J_{c3}(L)$  will get smaller with system size increasing.  The transition point $J_{c3}$ in the 2D limit can be evaluated from the transition point $J_{c3}(L)$, which is located at the peak postion of ground state energy function of $L\times L$ sites with respect to $J_2$. For example, for $24\times 24$ the phase transition occurs at  $J_{c3}$($L$=24) $=0.633$, seen from Fig.~\ref{fig:transitionPoint}(a), which shows the $J_2$ dependence of energies for each size. A linear fit of $J_{c3}(L)$ versus $1/L$ for $L=10-24$ gives $J_{c3}=0.609(1)$ for $L\rightarrow \infty$, shown in Fig.~\ref{fig:transitionPoint}(b). One can also use $L=12-24$ for fitting, giving an extrapolated value  $J_{c3}=0.6100(3)$.  

\begin{figure}[htbp]
 \centering
 \includegraphics[width=3.4in]{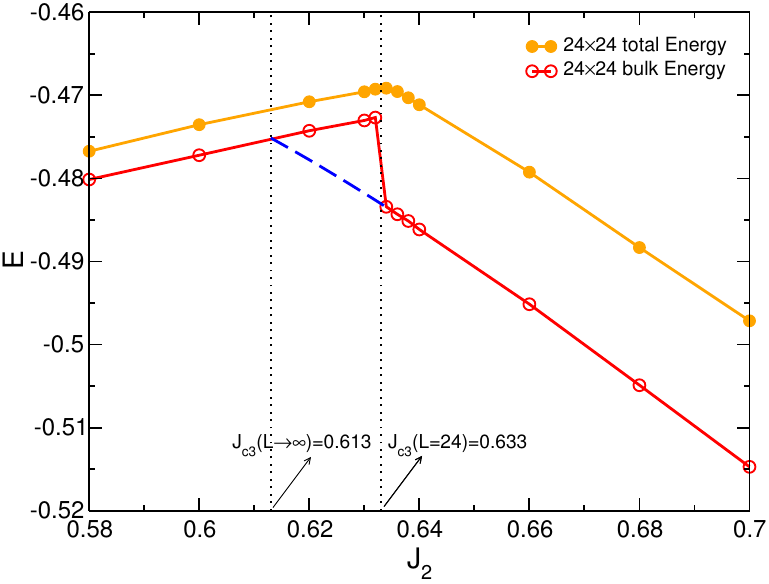}
 \caption{Another approach to estimate the first-order phase transition point $J_{c3}$. The blue dashed line is a third-order polynomial extrapolation from the bulk energies for $J_2>0.633$.
 }
 \label{fig:stripebulkEnergy}
 \end{figure}

Alternatively, the first-order phase transition point $J_{c3}$ in the 2D limit can be estimated only based on the $24\times 24$ open system. As we know, the energy in the 2D limit should be continuous with respect to $J_2$ at the phase transition. Generally, given $J_2$, the thermodynamic energy $E_{\infty}$ is almost the same as the deep bulk energy $E_b$ of an $L\times L$ system, i.e., $E_{\infty}=E_b(L)$, if $L$ is large enough, otherwise, finite size effects are still sizable in the bulk and $E_b(L)\ne E_{\infty}$. 
For example, as shown in Fig.\ref{fig:transitionPoint}(a), at $J_2=0.66$, $E_{\infty}=E_b(L)$ is not satisfied for $L=10$ whose first-order transition has not occured, but is true for $L\geq12$ which has experienced the phase transition. Generally, for $J_2\geq J_{c3}(L)$ and $J_2\leq J_{c3}$ it has $E_{\infty}=E_b(L)$, while for $J_{c3}\leq J_2\leq J_{c3}(L)$, $E_{\infty}\neq E_b(L)$. Now we use the relation between $E_{\infty}$ and $E_b(L)$ to estimate the first-order phase transition point $J_{c3}$ in the 2D limit.  We present the total bulk $L\times L$ energy  per site and the bulk $(L-8)\times (L-8)$ energy persite $E_b(L)$ in Fig.~\ref{fig:stripebulkEnergy}. We can see at the transition point $J_{c3}(L$$=$$24)$ $=0.633$, the bulk energy shows a sharp change,  which is discontinuous with respect to $J_2$, indicating $L=24$ is too small to be used to estimate $E_{\infty}$ for $J_{c3}\leq J_2\leq J_{c3}(L$=24). Since $E_{\infty}$ must be continuous w.r.t $J_2$, we can obtain the energies $E_{\infty}$ for $J_{c3} \leq  J_2 \leq  J_{c3}(L$=24) through the analytic continuation of the energies for $J_2 > J_{c3}(L$=$24)$.  Shown in Fig.~\ref{fig:stripebulkEnergy}, the analytic continuation is denoted by a blue dash line, thus the transition point in 2D limit is estimated at $J_{c3}\simeq0.613$, in good agreement with the above results.

 \section{Comparison of exponents}
\begin{figure}[htbp]
 \centering
 \includegraphics[width=3.4in]{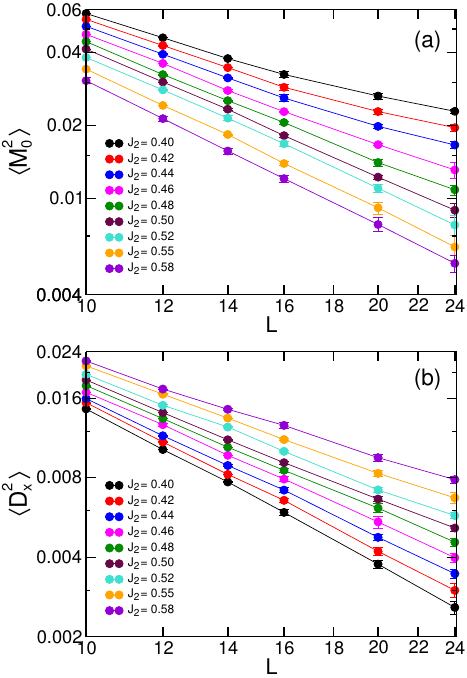}
 \caption{Log-log plot of  the AFM order $\langle {\bf M}^2_0 \rangle$  and of the dimer order $\langle D^2_x \rangle$ versus system size $L$ at different $J_2$.} 
 \label{fig:AFM_VBSexponent}
 \end{figure}

 Using the standard finite-size scaling formula~\cite{DQCP3,DQCP7}
 \begin{equation}
 O(L,J_2)=L^{-(z+\eta)}F_O[L^{1/\nu}(J_2-J_c)/J_c] \, ,
 \end{equation}
 where $O(L,J_2)$ is the spin order parameter (squared) $\langle {\bf M}^2_0 \rangle$ or the dimer order parameter (squared) $\langle D^2_x \rangle$, at the critical point $J_2=J_c$, one gets $O(L,J_c)\propto L^{-(1+\eta_{s,d})}$, assuming $z=1$. Therefore, we can compare the critical exponents $\eta_{s,d}$  from the data collapse with the exponents from the scalings $O(L,J_2)\propto L^{-(1+\eta'_{s,d})}$ at different $J_2$, to judge the correctness of the critical exponents.  It is expected that, when $J_2$ gets closer and closer to the critical point $J_c$ (within the critical phase), the exponent $\eta'_{s,d}$ converges to the critical exponents $\eta_{s,d}$.  In Fig.~\ref{fig:AFM_VBSexponent}, we present the log-log plot of $\langle {\bf M}^2_0 \rangle$  and $\langle D^2_x \rangle$ versus system size $L$. Exponents at different $J_2$ can be easily extracted, and they are listed in Table.~\ref{tab:ScalingExponents}. It shows that the critical exponents $\eta'_s$ and $\eta'_d$ in the critical QSL change continuously between the critical exponents  $\eta_s$ and $\eta_d$, obtained independently for data collapse.  Such a consistency demonstrates the reliability of obtaining critical exponents from data collapse.

\begin{figure}[htbp]
 \centering
 \includegraphics[width=3.4in]{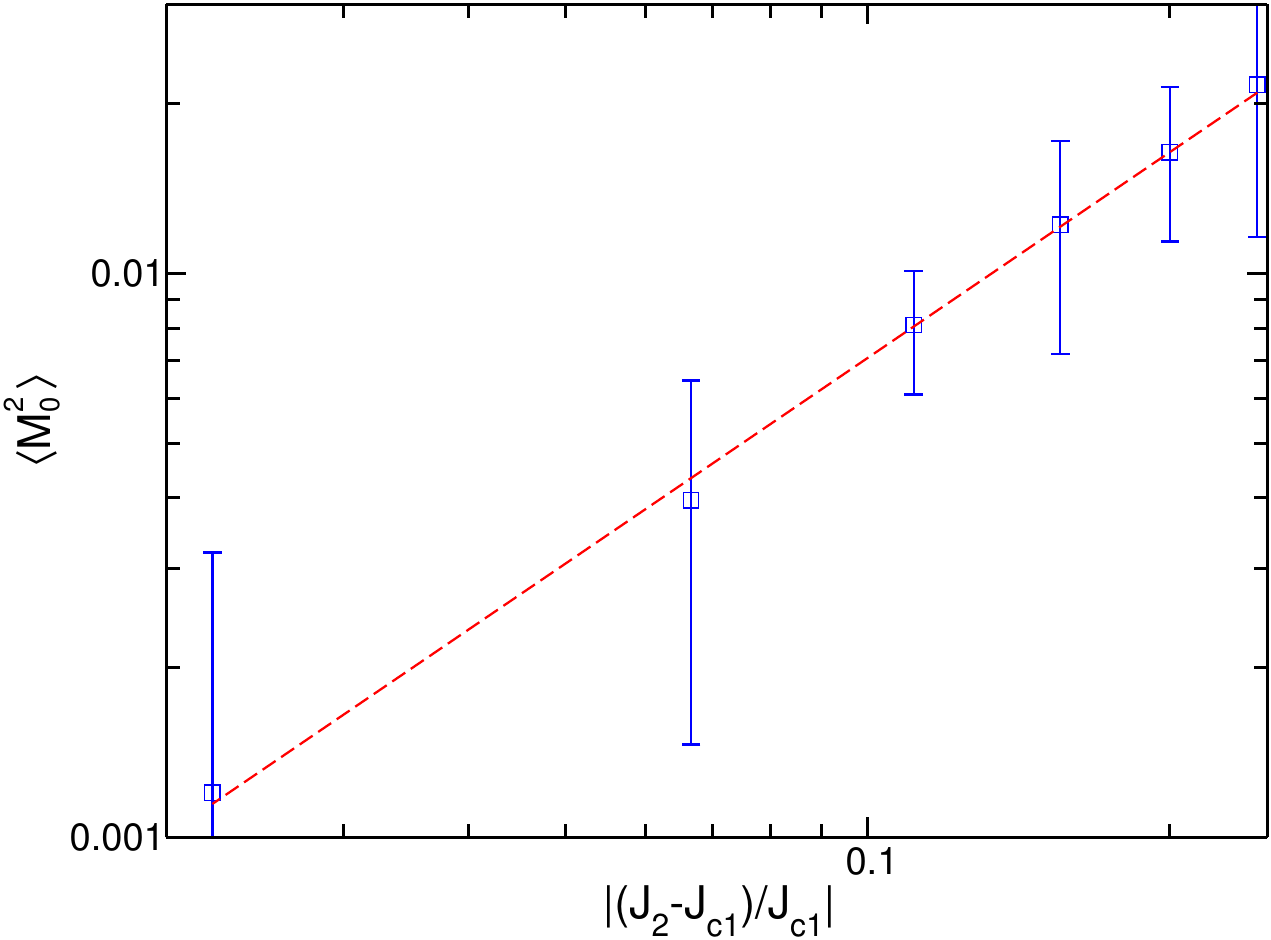}
 \caption{Estimating the $\beta$ exponent from the $L\rightarrow\infty$ extrapolated value of $\langle {\bf M}^2_0 \rangle$ plotted versus the distance from the critical value $J_{c1}=0.45$ in the interval from $J_2=0.34$ to $0.44$ (log-log plot).  } 
 \label{fig:betaExponent}
 \end{figure}

We can also fit the $\beta$ exponent of the antiferromagnetic order. An easy way to estimate $\beta$ is to use the relation $\beta/\nu=(1+\eta_{s1})/2$, which gives $\beta\approx 0.68$ by using $\eta_{s1}=0.38$ and $\nu=0.99$. We also try to esimate $\beta$ directly from $\langle {\bf M}^2_0\rangle \sim \Big(\frac{J-J_{c1}}{J_{c1}}\Big)^{2\beta}$ by using the 2D limit extrapolated magnetic order $\langle {\bf M}^2_0 \rangle$. Shown in Fig.~\ref{fig:betaExponent}, although the 2D $\langle {\bf M}^2_0\rangle$ values have large error bars, the log-log plot of  $\langle {\bf M}^2_0\rangle$ gives $\beta \approx 0.60$, consistent with the one from the equation $\beta/\nu=(1+\eta_{s1})/2$ .

   \begin{table*}[htbp]
   \centering
 \caption {Exponents obtained from the scalings $\langle M^2_0\rangle\propto L^{-(1+\eta'_s)}$ and $\langle D^2_x\rangle \propto L^{-(1+\eta'_d)}$ at different $J_2$ within the  critical QSL. The values corresponding to $J_{c1}$ and $J_{c2}$ are the critical exponents $\eta_s$ and $\eta_d$ obtained from data collapse.}
	\begin{tabular*}{\hsize}{@{}@{\extracolsep{\fill}}cccccccc@{}}
		\hline\hline
	      exponent &   $J_{c1}$     & 0.46 &0.48 & 0.50 & 0.52 & 0.55    & $J_{c2}$ \\ \hline
 		$\eta'_s$ & 0.38(3)     & 0.48(3) & 0.61(2)&0.76(1)&0.82(2) & 0.91(2)  & 0.96(4) \\
 		$\eta'_d$ & 0.72(4)    & 0.65(1) & 0.55(1)&0.49(3)&0.42(2)&0.32(2) & 0.26(3)   \\
 		\hline\hline
	\end{tabular*}
\label{tab:ScalingExponents}	
\end{table*}

\section{RVB on open boundary finite systems}
 The resonant valence bond state (RVB) is described by a $D=3$ PEPS with a single tensor and was used to investigate the ground state of the $J_1$-$J_2$ model at $J_2=0.5$. It is a gapless spin liquid state, which has been well studied on cylindrical  and periodic systems~\cite{WangRVB}. On open boundary conditions, we can deal with very large systems with extremely high precision, hence such a state provides an excellent benchmark to study the behavior of spin liquid state on open boundary  systems.

\begin{figure}[htbp]
 \centering
 \includegraphics[width=3.4in]{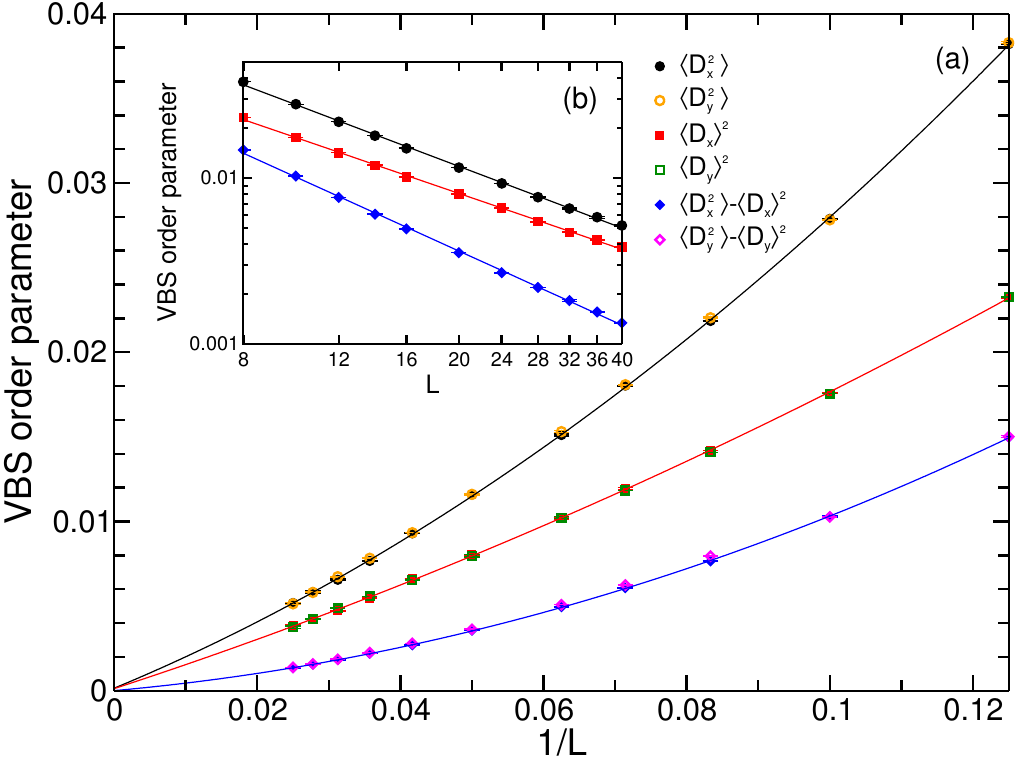}
 \caption{(a) Finite-size scaling of the VBS order parameter of the RVB  state on systems with open boundaries with up to $40\times40$ sites.  The extrapolations are performed through a second order polynomial fitting. (b) The log-log plot of order parameters versus system size $L$ . }
 \label{fig:RVBvbs}
 \end{figure}

First we consider order parameters on $L\times L$ systems up to $40\times 40$ sites. The spin AFM order decreases very rapidly (not shown) consistently with the short-range nature of the spin-spin correlations. Fig.~\ref{fig:RVBvbs}(a) shows the finite-size scaling of VBS order patameters, using two definitions, $\langle D^2_{\alpha} \rangle$ and $\langle D_{\alpha} \rangle^2$.
Both of them are zero in the 2D limit. Increasing system size $L$,  they decay as $L^{-1.23}$ and $L^{-1.12}$, respectively, as shown in Fig.~\ref{fig:RVBvbs}(b). Note that $\langle D^2_{\alpha} \rangle$ - based on the bond-bond correlations at all distances and $\langle D_{\alpha} \rangle^2$ - induced by the boundaries -  are both nonzero (and equal) in VBS states~\cite{zhao2020}, but zero in spin liquid states for $L\rightarrow \infty$. Therefore,  $\langle D^2_{\alpha} \rangle-\langle D_{\alpha} \rangle^2$ is always zero in the 2D limit in the VBS and is {\it not} a valid VBS order parameter on open boundary systems. 

 \begin{figure}[htbp]
 \centering
 \includegraphics[width=3.4in]{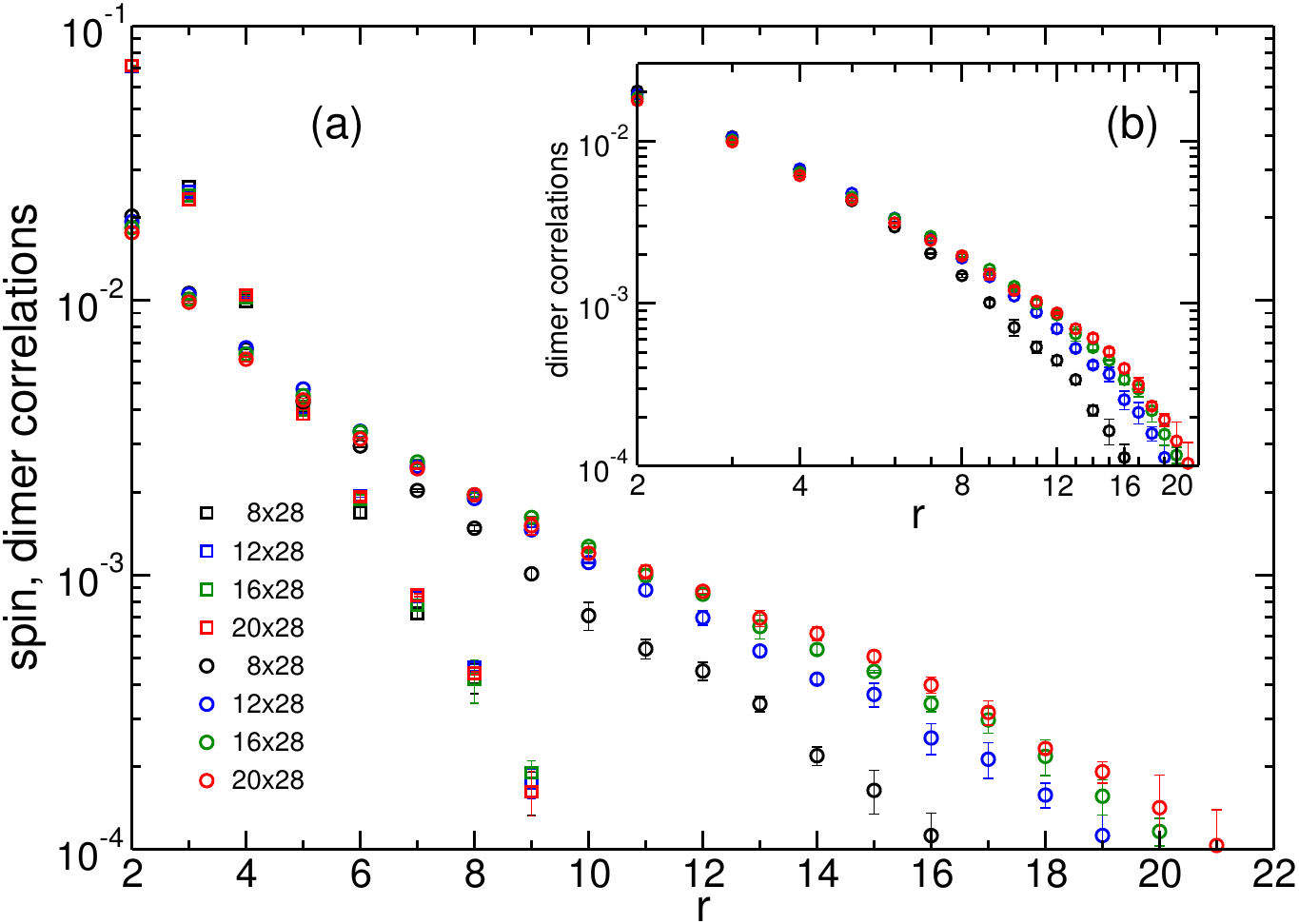}
 \caption{ Log-linear plot of spin (square) and dimer correlations (circle) of the RVB state on long strips along the central line $y=L_y/2$. The reference site is 3 lattice spacing away from the left edge. The inset  shows the dimer correlations on a log-log plot. }
 \label{fig:stripeRVBcorr}
 \end{figure}
  
As is studied in Ref.\cite{WangRVB} on cylindrical and periodic systems, the spin correlation functions of RVB decay exponentially, and its dimer correlation functions decay as a power law. Now we investigate the spin and dimer correlations on $L_y\times L_x$ strips to further understand the role of the boundaries, keeping $L_x=28$ fixed and varying $L_y$. Fig.~\ref{fig:stripeRVBcorr}(a) is a log-linear plot of the spin and dimer correlations with respect to the distance $r$.  The values of the spin correlations, at all distances, depend barely on the width of the strip, showing a clear exponential decay (with a rather short correlation length). The dimer correlations show a different  behavior. Increasing the system width $L_y$, we observe that short-distance correlations get smaller and converge but the long-distance ones get larger. So for large systems the dimer correlations will exhibit a long tail, indicating a power law decay. These are special features enabling to distinguish power law from exponential decays.  In Fig.~\ref{fig:stripeRVBcorr}(b), a log-log plot of the dimer correlations is also shown. It is notable that  the long-distance values bend down and show some deviation from a power law, which is just caused by the edge effects from the right open boundary.

   \section{Extracting decay length of hDOP}
 
     \begin{figure}[htbp]
 \centering
 \includegraphics[width=3.4in]{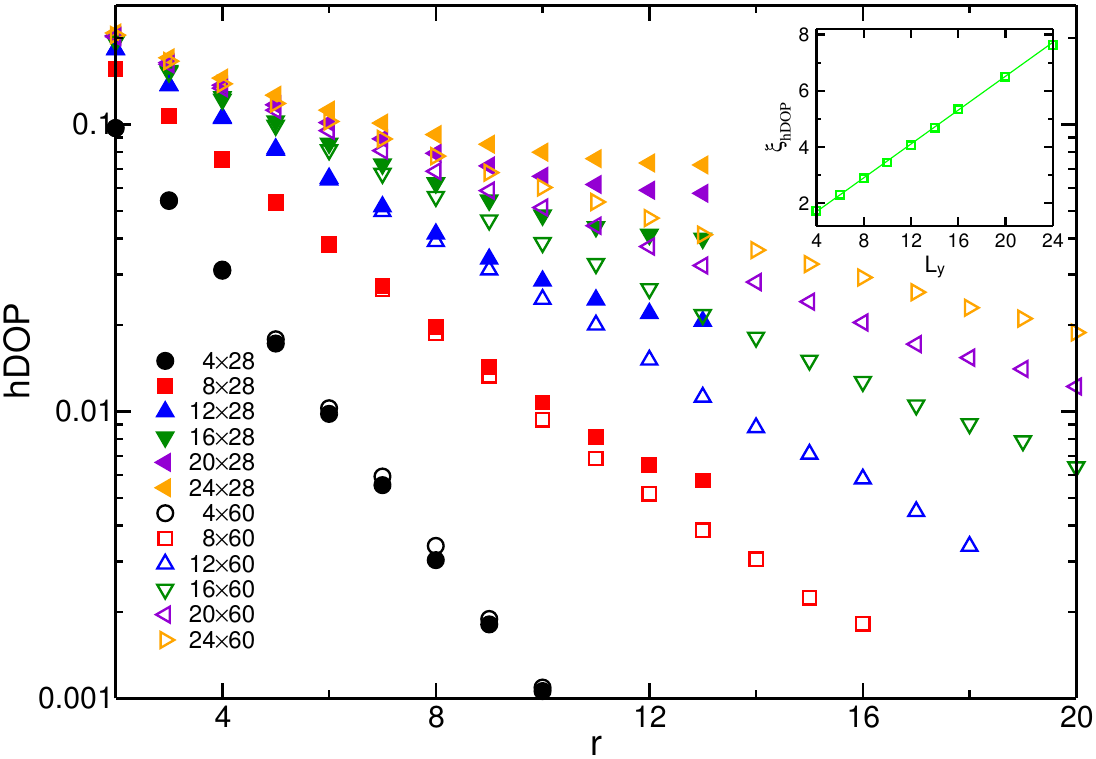}
 \caption{hDOP of the RVB state plotted as a function of the distance from the reference site along the central line $y=L_y/2$. Finite size effects on the decay length are controled by comparing results obtained for $L_x$ =28 and $60$. The inset shows a linear-linear plot of the hDOP decay length versus system width $L_y$.}
 \label{fig:BondHdopRVB}
 \end{figure}
 
\begin{figure}[htbp]
 \centering
 \includegraphics[width=3.4in]{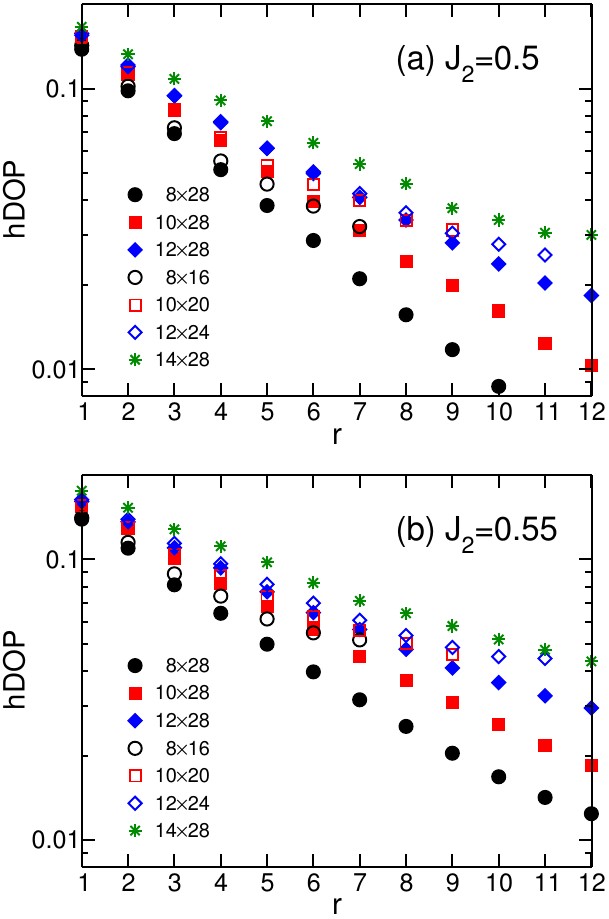}
 \caption{hDOP plotted as a function of the distance from the reference site along the central line $y=L_y/2$, for the $J_1$-$J_2$ model at $J_2=0.5$ (a) and $0.55$ (b), computed with a $D=8$ PEPS. Lengths $L_x=2L_y$ and  $L_x=28$ are considered, to control the effect of a finite $L_x$.}
 \label{fig:BondHdopJ050_055}
 \end{figure}

  The horizontal dimer order parameter (hDOP) is defined as the strong and weak energy bond difference (see main text). For long strips with a given system width $L_y$, the extracted decay length $\xi_{\rm hDOP}$ of the hDOP may be influenced by the system length $L_x$. We first compare the changes of the hDOP of the RVB state for $L_x=28$ and $L_x=60$, as shown in Fig.~\ref{fig:BondHdopRVB}. We can see that, for $L_y\geq 16$, one needs a relative large value of $L_x$ such as $L_x=60$ to minimize finite-size effects on $\xi_{\rm hDOP}$, while, for the other cases, $L_x=28$ is long enough to obtain the correct decay length.   For the $J_1$-$J_2$ model, we compare the hDOP on $L_x=28$ and $L_x=2L_y$, as shown in Fig.~\ref{fig:BondHdopJ050_055}. It can be seen that $L_x=28$ is long enough to extract $\xi_{\rm hDOP}$ by using the hDOP values from $r\leq 7$, for all cases.

 \section{Spin correlations at different $J_2$}

\begin{figure*}[htbp]
 \centering
 \includegraphics[width=6.8in]{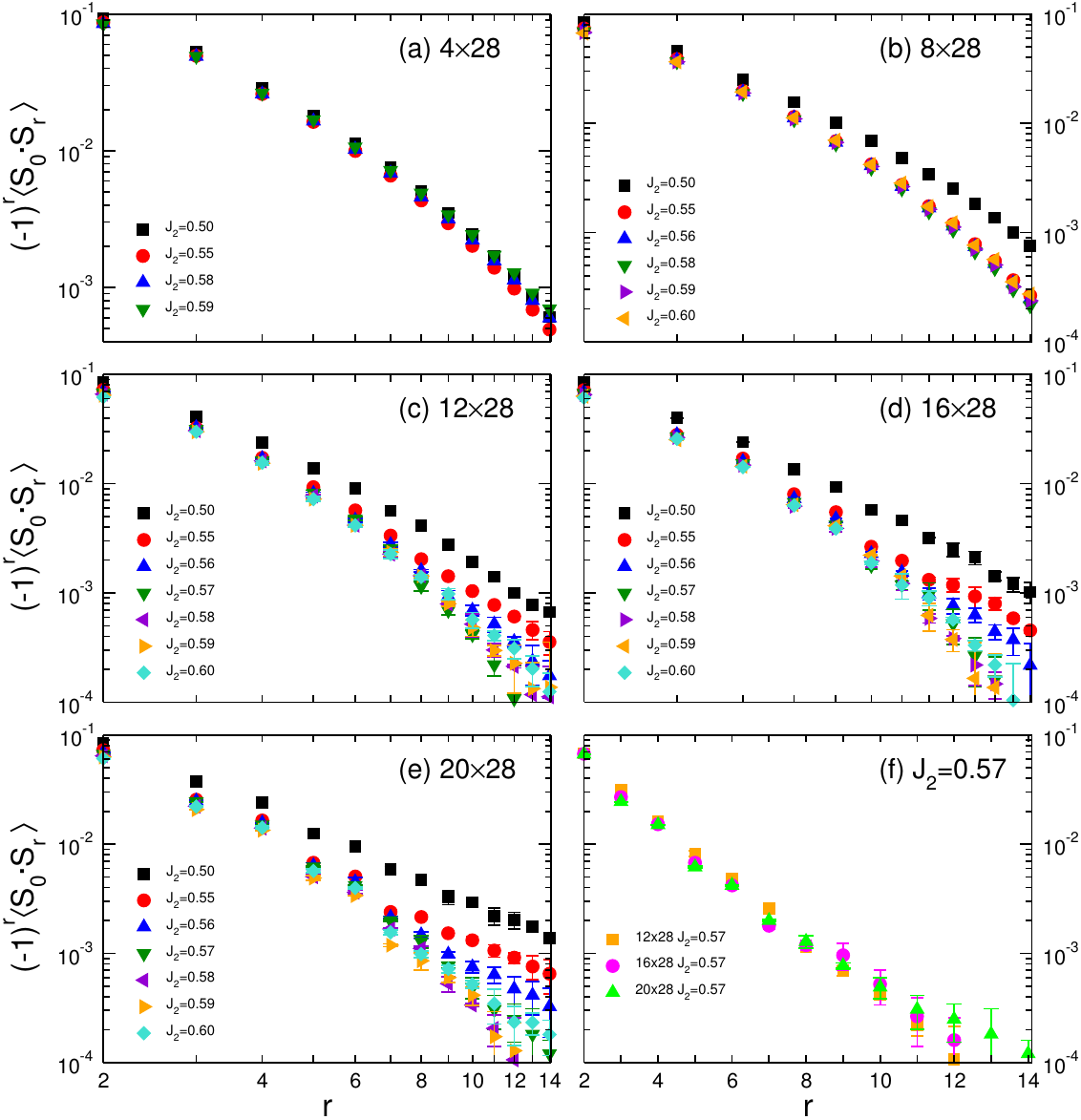}
 \caption{Log-log plot of spin-spin correlation functions at different $J_2$ ranging from 0.5 to 0.6 for $L_y\times L_x$ systems with $L_y=4$ (a), $L_y=8$ (b), $L_y=12$ (c), $L_y=16$ (d) and $L_y=20$ (e). As a comparison, we also use the semi-log plot for the spin-spin correlation functions at $J_2=0.57$ with different values of $L_y$ (f). We (very crudely) estimate the spin correlation length is $\xi\sim 2.2(3)$ for $L_y=20$.
 (a,b) have been obtained using  SU(2)-DMRG and (c,d,e,f)  using a $D=8$ PEPS.}
 \label{fig:AllSpinCorrJ2}
 \end{figure*}

 Here we study the finite-size effects on the spin correlations in the QSL and the VBS phases. Spin correlations obtained on long $L_y\times  28$ open strips, for different $J_2$ values, are shown in Fig.~\ref{fig:AllSpinCorrJ2}. In the narrowest strip, $L_y=4$, the correlations vary only slightly in the range  $J_2=0.5-0.6$. Increasing the width to $L_y=8$, the correlations at $J_2=0.5$ start to deviate from the other ones in the range $J_2=0.55-0.6$. Further increasing $L_y$ to 12, 16 and 20, we can see that the behaviors at $J_2=0.5$ and $0.55$, on one hand, and at $J_2=0.57-0.60$, on the other hand, deviates qualitatively, showing different decay behavior. Also data obtained for $L_y=12$, $16$ and $20$ are quantitatively quite close, indicating small remaining finite-size effects in contrast to $L_y=4$ and $8$. 
 In addition, note that, the spin correlations show very fast decay within the interval from $J_2=J_{c2}$ to 0.6, possibly fastest around $J_2=0.58$. indicating a minimum of the correlation length (or equivalently a maximum of the triplet gap) in the VBS phase.

\bibliography{j1j2paper_2021-10-23}

\end{document}